\documentclass[10pt]{iopart}
\usepackage[dvips,dvipdfm]{graphicx}
\usepackage{amssymb}
\usepackage{bm}
\renewcommand{\vec}[1]{\boldsymbol{#1}}

\begin{document}
\title
[Dynamics on finitely connected random graphs with arbitrary degree distributions]
{Parallel dynamics of disordered Ising spin systems on finitely connected directed random graphs with arbitrary degree distributions}
\author{Kazushi Mimura$^\dag$ and A.C.C. Coolen$^\ddag$}
\address{
  \dag \
  Faculty of Information Sciences, Hiroshima City University,
  3-4-1 Ohtsuka-Higashi, Asaminami-Ku, Hiroshima, 731-3194, Japan \\
  \ddag \
  Department of Mathematics, King's College London, The Strand,\\ London WC2R 2LS, UK
}
\ead{mimura@hiroshima-cu.ac.jp$,$ ton.coolen@kcl.ac.uk}

\begin{abstract}
We study the stochastic parallel dynamics of Ising spin systems
defined on finitely connected directed random graphs with arbitrary degree distributions,
using generating functional analysis.
For fully asymmetric graphs the dynamics of the system can be completely solved, due to the asymptotic absence of loops.
For arbitrary graph symmetry, we solve the dynamics exactly for the first few time steps, and we construct approximate stationary solutions.
\end{abstract}


\maketitle

\section{Introduction}

Problems defined on finitely connected (partially) random graphs
have been studied intensively in various fields of science.
Examples from physics are models of
spin glasses and related magnetic systems
\cite{Viana1985,Kanter1987,Mezard1987a,Mezard1987b,Mezard1987c,Mottishaw1987,Wong1988,Monasson1988,Coolen2005,PerezVicente2008}. 
A second field is information theory and computer science, 
where calculations involving finitely connected random graphs emerge in the context of error correcting codes 
\cite{Sourlas1989, Kabashima1999, Vicente1999, Murayama2000, Nakamura2001, Nishimori2001, Alamino2007, Neri2008},
lossy data compression \cite{Matsunaga2003, Murayama2003, Murayama2004, Wainwright2005, Ciliberti2005, Ciliberti2006, Martinian2006, Mimura2009a}, 
CDMA multi-user detection \cite{Yoshida2005, Raymond2007, Guo2008}, 
and combinatorial optimization \cite{Kirkpatric1994, Monasson1998a, Monasson1998b, Monasson1999, Zdeborova2008}. 
In biology and the social and economical sciences one finds processes on finitely connected random graphs 
in the context of neural \cite{Castillo2004a, Castillo2004b} as well as proteomic and gene regulation networks,
and small world models \cite{Gitterman2000, Nikoletopoulos2004, Hatchett2005b, Skantzos2005}. 
In all these fields, techniques of statistical mechanics have been decisive in making progress.
The initial focus in research has been on establishing  the equilibrium properties of processes on finitely connected random graphs, 
mainly by applying replica theory and the cavity method. 
In a second wave also the dynamical properties of processes on finitely connected random graphs have begun to be investigated in detail, see e.g. 
\cite{Semerjian2003, Semerjian2004a, Semerjian2004b, Hatchett2004, Hatchett2005a, Skantzos2007, Mozeika2008, Mozeika2009,Mimura2009b, Neri2009}. 
In real-world applications involving finitely connected random graphs, 
there are many situations in which it is essential for dynamical properties to be understood quantitatively, 
this is especially true for technological applications, 
such as the decoding of error correcting codes and the detection dynamics of CDMA multiuser detectors, 
and for biological information processing systems where the absence of detailed balance rules out equilibrium methods.
\par
In most of the problems described above, the dynamical variables and their interactions are represented 
as nodes and edges of finitely connected graphs, respectively. 
The simplest such graphs are Poissonian ones, i.e. sparse Erd\"os-R\'enyi graphs with a Poissonnian degree distribution in the thermodynamic limit.
However, the broad spectrum of problems to be understood demand dynamical techniques that are able to treat more general random graph ensembles. 
The dynamics of processes on graphs drawn from ensembles with arbitrary degree distributions have so far been studied 
using dynamical replica theory \cite{Hatchett2005a,Mozeika2009} and the cavity method \cite{Neri2009}.
One particular dynamical formalism, which enjoys the appeal of full exactness (if it applies),
is the generating functional method of De Dominicis \cite{Dominicis1978}. This method was first used by
Hatchett et al to investigate the parallel dynamics of bond-disordered Ising spin systems on finitely connected random graphs \cite{Hatchett2004},
but their study was limited to finitely connected Poissonian random graphs. 
In information theoretic problems and modeling real-world systems, 
there are many examples which are characterized by a finitely connected random graphs with a specific degree distributions. 
For example, a power-law distribution is often utilized to discuss 'small-world' networks. 
In the framework of error correcting codes, a one-point distribution and specific (complex) degree distributions which give more high performance are applied. 
\par
In this paper we generalize the analysis of \cite{Hatchett2004}: we use generating functional techniques 
to analyze the parallel stochastic dynamics of Ising spin models defined on finitely connected random graphs, 
which are drawn from ensembles in which the degree distributions can be chosen arbitrarily 
(including a power-low distribution, a one-point distribution and so on). 
We first derive general dynamical order parameter equations, which give a transparent exact description of collective processes in finitely connected Ising systems.
We then consider the simplest case of fully asymmetric finitely connected random graphs.
Here the theory acquires a simple form due to the absence of the loops, and the dynamics becomes very simple.
Away from fully asymmetric graphs, we calculate the first few time steps exactly, and construct (for fully symmetric graphs) approximate equilibrium  solutions.
To confirm the validity of our theory, we present numerical results for some typical conditions and typical graph ensembles; 
although our theory can handle arbitrary degree distributions, our examples are mostly regular random graphs (for simplicity).

\section{Model definitions}
\par
Let us consider a system of $N$ Ising-type spins, placed on the edges of a finitely connected directed random graph with an as yet arbitrary degree distribution.
The dynamics are given by a Markov process which represents synchronous stochastic spin alignment to local fields.
The probability $P_t(\vec{\sigma})$ of finding the microscopic state $\vec{\sigma}=(\sigma_1,\cdots,\sigma_N) \in \{-1,1\}^N$ at time $t$ can be written as
\begin{eqnarray}
  & & P_t(\vec{\sigma})
      = \sum_{\vec{\sigma}^\prime\in\{-1,1\}^N} W_{t-1}[\vec{\sigma}|\vec{\sigma}^\prime] P_{t-1}(\vec{\sigma}^\prime),
      \label{eq:process} \\
  & & W_{t-1}[\vec{\sigma}|\vec{\sigma}^\prime]
      = \prod_{i=1}^N \frac{e^{\beta \sigma_i h_i(\vec{\sigma}^\prime,t-1)}}{2 \cosh [\beta h_i(\vec{\sigma}^\prime,t-1)]}
\end{eqnarray}
where the $h_i(\vec{\sigma},t)$ are local fields, defined as
\begin{equation}
  h_i(\vec{\sigma},t) = \frac{1}{c} \sum_{j \ne i}^N c_{ij} J_{ij} \sigma_j + \theta_i(t).
\end{equation}
They involve a connectivity matrix $\vec{c}=\{c_{ij}\} \in \{0,1\}^{N \times N}$, whose entries specify which spins interact, 
and define a (generally directed) random graph. The parameters $\theta_i(t) \in \mathbb{R}$ are time dependent external fields, and $c>0$
The bond strengths $J_{ij}$ are symmetric, i.e., $J_{ij}=J_{ji}$, and are drawn independently from a bond distribution $\tilde{p}(J)$.
We define the degree $k_i$ of node $i$ (i.e. the number of links to the node $i$) as
\begin{equation}
  k_i
  = \sum_{j=1}^N c_{ij},
\end{equation}
(this would often be called the `in-degree'). 
The entries of the matrix $\vec{c}$ are chosen randomly according to the following connectivity distribution, in which all $N$ degrees $k_i$ are constrained,
\begin{equation}
  \hat{p}(\vec{c})
  = \frac
    {\displaystyle                \biggl( \prod_{i=1}^N \prod_{j>i}^N \hat{p}(c_{ij}) \hat{p}(c_{ji}|c_{ij}) \biggr) \biggl( \prod_{i=1}^N \delta_{k_i,\sum_{j=1}^N c_{ij}} \biggr)}
    {\displaystyle \sum_{\vec{c}^\prime} \biggl( \prod_{i=1}^N \prod_{j>i}^N \hat{p}(c^\prime_{ij}) \hat{p}(c^\prime_{ji}|c^\prime_{ij}) \biggr) \biggl( \prod_{i=1}^N \delta_{k_i,\sum_{j=1}^N c^\prime_{ij}} \biggr)},
  \label{eq:ConnectivityDistribution}
\end{equation}
where
\begin{eqnarray}
  & & \hat{p}(c_{ij})
      = \frac cN \delta_{c_{ij},1} + \biggl( 1 - \frac cN \biggr) \delta_{c_{ij},0}, \\
  & & \hat{p}(c_{ji}|c_{ij})
      = \varepsilon \delta_{c_{ji},c_{ij}} + (1-\varepsilon) \biggl[ \frac cN \delta_{c_{ji},1} + \biggl( 1 - \frac cN \biggr) \delta_{c_{ji},0} \biggr],
\end{eqnarray}
(with the Kronecker symbol $\delta_{mn}=1$ if $m=n$, and $\delta_{mn}=0$ otherwise).
All $k_i$ are randomly and independently drawn from a given degree distribution $p(k)$, with $\sum_k kp(k)=c$, so
 $c = \lim_{N\to\infty}\frac 1N \sum_{i=1}^N \sum_{j=1}^N c_{ij}$.
This procedure generates an ensemble of random graphs, with a prescribed degree distribution $p(k)$ and a
parameter $\varepsilon$ that controls the symmetry of the graph.
For $\varepsilon=1$, the connectivity becomes symmetric. In the remainder of this paper
we will write averages over the bond variables $\{J_{ij}\}$ as $\langle \cdots \rangle_{J} = \int\!\rmd J~ \tilde{p}(J) (\cdots)$
and averages over both the microscopic graph and bond variables, i.e., over $\{c_{ij}, J_{ij}\}$, as $\overline{[\cdots]}$.

\section{Generating functional analysis}

\subsection{The disorder-averaged generating functional}
\par
We follow the approach of \cite{Hatchett2004}, but apply it to the present generalized spin model.
We assume that the macroscopic behaviour of our system depends only on the statistical properties of the disorder.
The joint distribution of any trajectory $\vec{\sigma}(0),\cdots,\vec{\sigma}(t_m)$ is given
by products of the individual transition probabilities of the Markov chain:
\begin{equation}
  P[\vec{\sigma}(0),\cdots,\vec{\sigma}(t_m)]
  = p_0[\vec{\sigma}(0)] \prod_{t=0}^{t_m-1} W[\vec{\sigma}(t+1)|\vec{\sigma}(t)],
\end{equation}
where $p_0[\vec{\sigma}(0)]$ represents the initial conditions.
The generating functional $Z[\vec{\psi}]$ for the process  is now defined as \cite{Dominicis1978}
\begin{equation}
  \overline{Z[\vec{\psi}]}
  = \overline{\biggl\langle \exp \biggl[ -\rmi \sum_{i=1}^N \sum_{t=0}^{t_m} \psi_i(t) \sigma_i(t) \biggr] \biggr\rangle}
\end{equation}
where $\langle \cdots \rangle$ denotes averaging over the microscopic process, viz.
\begin{equation}
  \langle \cdots \rangle
  = \sum_{\vec{\sigma}(0) \in \{-1,1\}^N}\!\! \cdots \!\!\sum_{\vec{\sigma}(t_m) \in \{-1,1\}^N}
    p[\vec{\sigma}(0),\cdots,\vec{\sigma}(t_m)] (\cdots),
\end{equation}
and $\vec{\psi}=(\psi_i(t))$ denote the generating fields.
We isolate the local fields at all stages by inserting appropriate integrals over integral representations of the Dirac delta function:
\begin{equation}
\hspace*{-10mm}
  1 = \int\! \{ \rmd\vec{h} \rmd\hat{\vec{h}} \}
      \prod_{i=1}^N \prod_{t=0}^{t_m\!-1}
      \exp \biggl[ \rmi \hat{h}_i(t) \biggl( h_i(t) - \frac{1}{c} \sum_{j \ne i}^N c_{ij} J_{ij} \sigma_j(t) - \theta_i(t) \biggr) \biggr],
\end{equation}
where $\{ \rmd\vec{h} \rmd\hat{\vec{h}} \} = \prod_{i=1}^N \prod_{t=0}^{t_m-1} [\rmd h_i(t)\rmd\hat{h}_i(t)/2\pi]$.
The generating functional then takes the form
\begin{eqnarray}
\hspace*{-15mm}  \overline{Z[\vec{\psi}]}
  &=& \int\! \{ \rmd\vec{h} \rmd\hat{\vec{h}} \}
      \sum_{\vec{\sigma}(0)} \!\!\cdots\!\! \sum_{\vec{\sigma}(t_m)} p_0[\vec{\sigma}(0)] \;
      \overline{ \exp \biggl[ - \frac{\rmi}{c} \sum_{i=1}^N \sum_{t=0}^{t_m-1} \hat{h}_i(t) \sum_{j \ne i}^N c_{ij} J_{ij} \sigma_j(t) \biggr] } \nonumber \\
      \hspace*{-15mm}
  & & \times \prod_{i=1}^N \prod_{t=0}^{t_m-1}
      \rme^{ \rmi \hat{h}_i(t) [ h_i(t) - \theta_i(t) ]
      - \rmi \psi_i(t) \sigma_i(t)
      + \beta \sigma_i(t+1) h_i(t)
      - \ln 2 \cosh [ \beta h_i(t) ]}.
  \label{eq:AveragedZ}
\end{eqnarray}
Hereafter we will change our notation from $\vec{\sigma}(t)=(\sigma_1(t),\cdots,\sigma_N(t))$ (the $N$-spin configuration at time $t$) 
to $\vec{\sigma}_i=(\sigma_i(0),\cdots,\sigma_i(t_m))$ (the path taken from $t=0$ to $t=t_m$ by spin $i$). 
Similarly we define single site paths for external, local and conjugate fields, viz. 
$\vec{\theta}_i=(\theta_i(0),\cdots,\theta_i(t_m))$, $\hat{\vec{h}}_i=(\hat{h}_i(0),\cdots,\hat{h}_i(t_m))$, etc.
\par
Using the integral form of the Kronecker delta to represent the degree constraints,
\begin{eqnarray}
   \delta_{k_i,\sum_{j=1}^N c_{ij}} = \int_0^{2\pi}\! \frac{\rmd\omega_i}{2\pi} \exp \biggl[ \rmi \omega_i \biggl( k_i - \sum_{j=1}^N c_{ij} \biggr) \biggr]
\end{eqnarray}
the term in (\ref{eq:AveragedZ}) containing the disorder $\{c_{ij}, J_{ij}\}$ becomes
\begin{eqnarray}
  & &\hspace*{-20mm}  \overline{ \exp \biggl[ - \frac{\rmi}{c} \sum_{i=1}^N \sum_{t=0}^{t_m-1} \hat{h}_i(t) \sum_{j \ne i}^N c_{ij} J_{ij} \sigma_j(t) \biggr] } \nonumber \\
  &=& \frac1{Z_c}
      \biggl( \prod_{i=1}^N \int\! \frac{\rmd\omega_i}{2\pi} \rme^{\rmi \omega_i k_i} \biggr)
      \exp \biggl[ \frac c{2N} \sum_{i=1}^N \sum_{j=1}^N \biggl\langle
         \varepsilon    \rme^{ -\frac{\rmi J}{c} ( \vec{\sigma}_i \cdot \hat{\vec{h}}_j + \vec{\sigma}_j \cdot \hat{\vec{h}}_i ) - \rmi ( \omega_i + \omega_j )} \nonumber \\
  & & + (1-\varepsilon) \rme^{ -\frac{\rmi J}{c}   \vec{\sigma}_i \cdot \hat{\vec{h}}_j - \rmi \omega_j }
      + (1-\varepsilon) \rme^{ -\frac{\rmi J}{c}   \vec{\sigma}_j \cdot \hat{\vec{h}}_i - \rmi \omega_i }
      + \varepsilon - 2 \biggr\rangle_{\!J} \biggr],
  \label{eq:DisorderTerm}
\end{eqnarray}
where $Z_c$ denotes the normalization constant of the connectivity distribution $\hat{p}(\vec{c})$ in (\ref{eq:ConnectivityDistribution}), 
i.e. $Z_c = \sum_{\vec{c}}( \prod_{i=1}^N \prod_{j>i}^N \hat{p}(c_{ij}) \hat{p}(c_{ji}|c_{ij}) ) ( \prod_{i=1}^N \delta_{k_i,\sum_{j=1}^N c_{ij}} )$. 
Details on the calculation of (\ref{eq:DisorderTerm}) can be found in \ref{app:disorder_average}. 
To achieve site factorization we choose factorized homogeneous initial conditions, i.e. $p_0[\vec{\sigma}(0)]=\prod_i p_{0}[\sigma_i(0)]$, a
nd we introduce the following order parameter functions
\begin{eqnarray}
  & & P(\vec{\sigma},\hat{\vec{h}}) = \frac 1N \sum_{i=1}^N \delta_{\vec{\sigma},\vec{\sigma}_i} \delta(\hat{\vec{h}}-\hat{\vec{h}}_i), \label{eq:DefP} \\
  & & Q(\vec{\sigma},\hat{\vec{h}}) = \frac 1N \sum_{i=1}^N \delta_{\vec{\sigma},\vec{\sigma}_i} \delta(\hat{\vec{h}}-\hat{\vec{h}}_i) \rme^{- \rmi \omega_i}. \label{eq:DefQ}
\end{eqnarray}
It should be noted that one could also have used the order parameter function
$R(\vec{\sigma},\hat{\vec{h}},\omega) = \frac 1N \sum_{i=1}^N \delta_{\vec{\sigma},\vec{\sigma}_i} \delta(\hat{\vec{h}}-\hat{\vec{h}}_i) \delta(\omega - \omega_i)$ 
to achieve site factorization in the generating functional. 
The reason for us to work with (\ref{eq:DefP}) and (\ref{eq:DefQ}) is that the latter will have transparent physical interpretations.
We determinne the initial state $\sigma_i(0)$ according to the distribution $p_0[\sigma_i(0)]=\frac12[1+\sigma_i(0)m(0)]$, where $m(0)$ denotes the initial magnetization. 
\par
To introduce the two order parameter functions in the generating functional,
we use the integral form of Dirac's delta function and insert
\begin{eqnarray}
 1 &=& \int\! \{\rmd P\rmd\hat{P}\} \exp \biggl[ \rmi N \sum_{\vec{\sigma}} \int\! \rmd\hat{\vec{h}}~ \hat{P}(\vec{\sigma},\hat{\vec{h}}) \nonumber \\
   & & \times \biggl( P(\vec{\sigma},\hat{\vec{h}}) - \frac 1N \sum_{i=1}^N \delta_{\vec{\sigma},\vec{\sigma}_i} \delta(\hat{\vec{h}}-\hat{\vec{h}}_i) \biggr) \biggr], \\
 1 &=& \int\! \{\rmd Q\rmd\hat{Q}\} \exp \biggl[ \rmi N \sum_{\vec{\sigma}} \int\! \rmd\hat{\vec{h}}~ \hat{Q}(\vec{\sigma},\hat{\vec{h}}) \nonumber \\
   & & \times \biggl( Q(\vec{\sigma},\hat{\vec{h}}) - \frac 1N \sum_{i=1}^N \delta_{\vec{\sigma},\vec{\sigma}_i} \delta(\hat{\vec{h}}-\hat{\vec{h}}_i) \rme^{- \rmi\omega_i} \biggr) \biggr],
\end{eqnarray}
where $\{\rmd P\rmd\hat{P}\}= \prod_{\vec{\sigma},\hat{\vec{h}}}[\rmd P(\vec{\sigma},\hat{\vec{h}})\rmd\hat{P}(\vec{\sigma},\hat{\vec{h}})N/2\pi]$,
and we choose the equivalent definition for the short-hand $\{\rmd Q\rmd\hat{Q}\}$.
We then have
\begin{equation}
  \overline{Z[\vec{\psi}]}
  = \frac 1{Z_c} \int \{\rmd P\rmd\hat{P}\rmd Q\rmd\hat{Q}\} \rme^{N \Psi[\{P,\hat{P},Q,\hat{Q}\}]},
  \label{eq:AveragedZ2}
\end{equation}
where
\begin{eqnarray}
  & & \hspace*{-10mm}
  \Psi[\{P,\hat{P},Q,\hat{Q}\}]
      = \frac{c(\varepsilon\! -\!2)}2
      + \frac c2 \sum_{\vec{\sigma},\vec{\sigma}^\prime} \int\! \rmd\hat{\vec{h}} \rmd\hat{\vec{h}}^\prime ~ A(\vec{\sigma},\hat{\vec{h}};\vec{\sigma}^\prime,\hat{\vec{h}}^\prime)
       + L[\{\hat{P},\hat{Q},\vec{\theta}\}]
      \nonumber \\
  & & \quad
      + \rmi \sum_{\vec{\sigma}} \int\! \rmd\hat{\vec{h}}~ \hat{P}(\vec{\sigma},\hat{\vec{h}}) P(\vec{\sigma},\hat{\vec{h}})
      + \rmi \sum_{\vec{\sigma}} \int\! \rmd\hat{\vec{h}}~ \hat{Q}(\vec{\sigma},\hat{\vec{h}}) Q(\vec{\sigma},\hat{\vec{h}})
\end{eqnarray}
with
\begin{eqnarray}
   A(\vec{\sigma},\hat{\vec{h}};\vec{\sigma}^\prime,\hat{\vec{h}}^\prime)
   &=&   Q(\vec{\sigma},\hat{\vec{h}}) Q(\vec{\sigma}^\prime,\hat{\vec{h}}^\prime)
         \langle \varepsilon \rme^{ -\frac{\rmi J}{c}(\vec{\sigma} \cdot \hat{\vec{h}}^\prime + \vec{\sigma}^\prime \cdot \hat{\vec{h}}) } \rangle_J \nonumber
         \\
   & & + P(\vec{\sigma},\hat{\vec{h}}) Q(\vec{\sigma}^\prime,\hat{\vec{h}}^\prime)
         \langle (1-\varepsilon) \rme^{ -\frac{\rmi J}{c}\vec{\sigma} \cdot \hat{\vec{h}}^\prime } \rangle_J
         \nonumber
         \\
   & & + Q(\vec{\sigma},\hat{\vec{h}}) P(\vec{\sigma}^\prime,\hat{\vec{h}}^\prime)
         \langle (1-\varepsilon) \rme^{ -\frac{\rmi J}{c}\vec{\sigma}^\prime \cdot \hat{\vec{h}} } \rangle_J
\end{eqnarray}
and
\begin{eqnarray}
  L[\{\hat{P},\hat{Q},\vec{\theta}\}]
  &=& \frac 1N \sum_{i=1}^N \ln \sum_{\vec{\sigma}} p_{0}[\sigma(0)]
      \nonumber
      \\
  & & \times \int\! \biggl( \prod_{t=0}^{t_m-1} \frac{\rmd h(t)\rmd\hat{h}(t)}{2\pi} \frac {\rme^{\rmi\hat{h}(t)[h(t)-\theta_i(t)]+\beta \sigma(t+1) h(t)}}{2 \cosh [\beta h(t)]} \biggr) \nonumber
      \\
  & & \qquad \times \rme^{-\rmi\hat{P}(\vec{\sigma},\hat{\vec{h}})} \frac{(-\rmi)^{k_i}}{k_i!} [\hat{Q}(\vec{\sigma},\hat{\vec{h}})]^{k_i}.
  \label{eq:Li}
\end{eqnarray}
In the above expressions we have neglected those terms that will vanish for $N \to \infty$,  and we have removed the now redundant generating fields $\vec{\psi}$.
Upon applying the law of large numbers, the term $L[\{\hat{P},\hat{Q},\vec{\theta}\}]$ simplifies to
\begin{eqnarray}
L[\{\hat{P},\hat{Q},\vec{\theta}\}] &=& \sum_{k=0}^\infty p(k) \ln \sum_{\vec{\sigma}} p_0[\sigma(0)]
\nonumber
\\&&\times
        \int\! \biggl( \prod_{t=0}^{t_m-1} \frac{\rmd h(t)\rmd\hat{h}(t)}{2\pi} \frac {\rme^{\rmi\hat{h}(t)[h(t)-\theta(t)]+\beta\sigma(t+1)h(t)}}{2 \cosh [\beta h(t)]} \biggr) \nonumber \\
  & & \qquad \times \rme^{-\rmi\hat{P}(\vec{\sigma},\hat{\vec{h}})} \frac{(-\rmi)^k}{k!} [\hat{Q}(\vec{\sigma},\hat{\vec{h}})]^k.
  \label{eq:L}
\end{eqnarray}
Functional extremization of $\Psi[\{P,\hat{P},Q,\hat{Q}\}]$ with respect to the kernels $P$, $\hat{P}$, $Q$ and $\hat{Q}$,
i.e. working out the equations $\delta \Psi/\delta P=\delta \Psi/\delta \hat{P}=\delta \Psi/\delta Q=\delta \Psi/\delta \hat{Q}=0$,
gives the following functional saddle-point equations:
\begin{eqnarray}
\hat{P}(\vec{\sigma},\hat{\vec{h}})
      &=& \rmi c \sum_{\vec{\sigma}^\prime} \int\! \rmd\hat{\vec{h}}^\prime~ Q(\vec{\sigma}^\prime,\hat{\vec{h}}^\prime)
      \langle (1-\varepsilon) \rme^{-\frac{\rmi J}{c} \vec{\sigma} \cdot \hat{\vec{h}}^\prime } \rangle_J , \\
P(\vec{\sigma}^\prime,\hat{\vec{h}}^\prime)
      &=& \sum_{k=0}^\infty p(k) \langle \delta_{\vec{\sigma},\vec{\sigma}'} \delta(\hat{\vec{h}} - \hat{\vec{h}}') \rangle_{\vec{\theta},k} , \\
\hat{Q}(\vec{\sigma},\hat{\vec{h}})
      &=& \rmi c \sum_{\vec{\sigma}^\prime} \int\! \rmd\hat{\vec{h}}^\prime~ P(\vec{\sigma}^\prime,\hat{\vec{h}}^\prime)
      \langle (1-\varepsilon) \rme^{-\frac{\rmi J}{c} \vec{\sigma}^\prime \cdot \hat{\vec{h}}  } \rangle_J \nonumber \\
  & & \qquad
      + \rmi c \sum_{\vec{\sigma}^\prime} \int\! \rmd\hat{\vec{h}}^\prime~ Q(\vec{\sigma}^\prime,\hat{\vec{h}}^\prime)
      \langle \varepsilon  \rme^{-\frac{\rmi J}{c} (\vec{\sigma} \cdot \hat{\vec{h}}^\prime + \vec{\sigma}^\prime \cdot \hat{\vec{h}}) } \rangle_J , \\
  Q(\vec{\sigma}^\prime,\hat{\vec{h}}^\prime)
      &=& \sum_{k=1}^\infty k p(k)
      \biggl\langle \frac{ \delta_{\vec{\sigma},\vec{\sigma}^\prime} \delta(\hat{\vec{h}} - \hat{\vec{h}}^\prime) }
                         { -\rmi \hat{Q}(\vec{\sigma},\hat{\vec{h}}) }
      \biggr\rangle_{\!\vec{\theta},k} ,
\end{eqnarray}
with a measure $\langle \cdots \rangle_{\vec{\theta},k}$, which is defined as
\begin{eqnarray}
  & & \langle f(\vec{\sigma},\hat{\vec{h}}) \rangle_{\vec{\theta},k}
      = \frac {\sum_{\vec{\sigma}} \int\! \rmd\hat{\vec{h}}~ f(\vec{\sigma},\hat{\vec{h}}) M_k(\vec{\sigma},\hat{\vec{h}}|\vec{\theta})}
              {\sum_{\vec{\sigma}} \int\! \rmd\hat{\vec{h}} M_k(\vec{\sigma},\hat{\vec{h}}|\vec{\theta})},
      \label{eq:measure} \\
  & & M_k(\vec{\sigma},\hat{\vec{h}}|\vec{\theta})
      = \rme^{-\rmi\hat{P}(\vec{\sigma},\hat{\vec{h}})}
        [-\rmi\hat{Q}(\vec{\sigma},\hat{\vec{h}})]^k
        p_0[\sigma(0)] \nonumber \\
  & & \qquad \qquad \qquad \times \prod_{t=0}^{t_m-1} \int\! \frac{\rmd h(t)}{2\pi} \frac{\rme^{\rmi\hat{h}(t)[h(t)-\theta(t)]+\beta\sigma(t+1)h(t)}}{2\cosh[\beta h(t)]}.
\end{eqnarray}
Performing an inverse Fourier transformation of $P(\vec{\sigma},\hat{\vec{h}})$ and $Q(\vec{\sigma},\hat{\vec{h}})$ gives, respectively,
\begin{eqnarray}
  & & P(\vec{\sigma}|\vec{\theta}^\prime)
      \equiv \int\! \rmd\hat{\vec{h}}~ \rme^{-\rmi\vec{\theta}^\prime\cdot\hat{\vec{h}}} P(\vec{\sigma},\hat{\vec{h}})
      = \sum_{k=0}^\infty p(k)
        \langle \delta_{\vec{\sigma},\vec{\sigma}^\prime}
        \rangle_{\vec{\theta}+\vec{\theta}^\prime,k},
      \label{eq:P} \\
  & & Q(\vec{\sigma}|\vec{\theta}^\prime)
      \equiv \int\! \rmd\hat{\vec{h}}~ \rme^{-\rmi\vec{\theta}^\prime\cdot\hat{\vec{h}}} Q(\vec{\sigma},\hat{\vec{h}})
      = \sum_{k=1}^\infty k p(k)
        \biggl\langle \frac{ \delta_{\vec{\sigma},\vec{\sigma}^\prime}}
                           {-\rmi \hat{Q}(\vec{\sigma},\hat{\vec{h}})   }
        \biggr\rangle_{\vec{\theta}+\vec{\theta}^\prime,k},
      \label{eq:Q}
\end{eqnarray}
The order parameter $P(\vec{\sigma}|\vec{\theta}^\prime)$ is the disorder-averaged probability of finding a single-spin trajectory $\vec{\sigma}$ 
with modified external fields $\vec{\theta}+\vec{\theta}^\prime$ (consisting of the original external fields $\vec{\theta}$ and the supplement $\vec{\theta}^\prime)$. 
We will consider the meaning of the order parameter $Q(\vec{\sigma}|\vec{\theta}^\prime)$ later. 
The present  situation is similar to that in the related studies \cite{Hatchett2004, Semerjian2003}. 
In terms of $P(\vec{\sigma}|\vec{\theta}^\prime)$ and $Q(\vec{\sigma}|\vec{\theta}^\prime)$, 
the order parameter functions $\hat{P}(\vec{\sigma},\hat{\vec{h}})$ and $\hat{Q}(\vec{\sigma},\hat{\vec{h}})$ can be rewritten as 
\begin{eqnarray}
  & & \hat{P}(\vec{\sigma},\hat{\vec{h}})
      = \rmi c (1-\varepsilon)
      \label{eq:P^} \\
  & & \hat{Q}(\vec{\sigma},\hat{\vec{h}})
      = \rmi c (1-\varepsilon) \sum_{\vec{\sigma}^\prime}
        \langle \rme^{-\frac{\rmi J}{c} \vec{\sigma}^\prime \cdot \hat{\vec{h}}} P(\vec{\sigma}^\prime|\vec{0}) \rangle_J \nonumber \\
  & & \qquad \qquad
      + \rmi c \varepsilon \sum_{\vec{\sigma}^\prime}
        \langle   \rme^{-\frac{\rmi J}{c} \vec{\sigma}^\prime \cdot \hat{\vec{h}}} Q(\vec{\sigma}^\prime|\frac{J}{c}\vec{\sigma}) \rangle_J.
      \label{eq:Q^}
\end{eqnarray}
Here we have also used the fact that $P(\vec{\sigma}|\vec{\theta}^\prime)$ and $Q(\vec{\sigma}|\vec{\theta}^\prime)$ are both normalized with respect to summation over $\vec{\sigma}$, a property which is demonstrated in the next section.

After substituting (\ref{eq:P^}, \ref{eq:Q^}) into (\ref{eq:P}, \ref{eq:Q}) and integrating over $\hat{\vec{h}}$,
we finally arrive at the following compact closed-form equations:
\begin{eqnarray}
\hspace*{-20mm}
  P(\vec{\sigma}|\vec{\theta}^\prime)
  &=& \sum_{k=0}^\infty p(k)
   \biggl( \prod_{\ell=1}^k  \int\! \rmd J_\ell~ \tilde{p}(J_\ell) \sum_{\vec{\sigma}_\ell} \biggl[
           \varepsilon  Q( \vec{\sigma}_\ell | \frac{1}{c} J_\ell \vec{\sigma} )
      + (1\!-\!\varepsilon) P( \vec{\sigma}_\ell | \vec{0}            ) \biggr] \biggr) \nonumber \\
      \hspace*{-20mm}
  & & \times p_0 [\sigma(0)] \prod_{t=0}^{t_m-1}
      \frac{ \exp( \beta \sigma(t+1) [ \theta(t) + \theta^\prime(t) + \frac{1}{c} \sum_{\ell=1}^k J_\ell \sigma_\ell(t) ] ) }
           { 2 \cosh(\beta [ \theta(t) + \theta^\prime(t) + \frac{1}{c}\sum_{\ell=1}^k J_\ell \sigma_\ell(t) ] ) }
             \label{eq:cP}
\\
\hspace*{-20mm}
  Q(\vec{\sigma}|\vec{\theta}^\prime)
  &=& \sum_{k=0}^\infty \frac{k\!+\!1}{c} p(k\!+\!1)
 \biggl( \prod_{\ell=1}^k  \int\! \rmd J_\ell~ \tilde{p}(J_\ell) \sum_{\vec{\sigma}_\ell} \biggl[
           \varepsilon  Q( \vec{\sigma}_\ell | \frac{1}{c} J_\ell \vec{\sigma} )
      + (1\!-\!\varepsilon) P( \vec{\sigma}_\ell | \vec{0}            ) \biggr] \biggr) \nonumber \\
      \hspace*{-20mm}
  & & \times p_0 [\sigma(0)] \prod_{t=0}^{t_m-1}
      \frac{ \exp( \beta \sigma(t+1) [ \theta(t) + \theta^\prime(t) + \frac{1}{c} \sum_{\ell=1}^k J_\ell \sigma_\ell(t) ] ) }
           { 2 \cosh(\beta [ \theta(t) + \theta^\prime(t) + \frac{1}{c} \sum_{\ell=1}^k J_\ell \sigma_\ell(t) ] ) }.
  \label{eq:cQ}
\end{eqnarray}
The equation for the order parameter $P(\vec{\sigma}|\vec{\theta}^\prime)$ contains the degree distribution in its bare form $p(k)$.
The equation for $Q(\vec{\sigma}|\vec{\theta}^\prime)$, on the other hand, involves the deformed degree measure $\frac{k+1}{c} p(k+1)$.
This is the only difference between the two expressions.
These above two equations are closed and exact and completely general; they cannot be simplified further 
without making specific parameter choices or assumptions.
\par 
In the case of Poissonian graph \cite{Hatchett2004}, the result looks like being represented in terms of only the local field distribution. 
However, in general both the local field and the cavity field are needed to represent the macroscopic dynamics. 
In the next section, we discuss the meaning of equations (\ref{eq:cP}) and (\ref{eq:cQ}) in detail. 

\subsection{Physical meaning of order parameter functions and interpretation of closed-form equations}

Starting from (\ref{eq:DefP}) and (\ref{eq:DefQ}) and using manipulations similar to those used in deriving (\ref{eq:Li}),
we can infer the physical meaning of $P(\vec{\sigma}|\vec{\theta}^\prime)$ and $Q(\vec{\sigma}|\vec{\theta}^\prime)$ at the saddle-point.
This gives
\begin{eqnarray}
  & & P(\vec{\sigma}|\vec{\theta}^\prime) |_{\rm saddle}
      = \langle P(\vec{\sigma}|\vec{\theta}^\prime) \rangle_*
      = \frac 1N \sum_{i=1}^N \biggl. \langle \delta_{\vec{\sigma},\vec{\sigma}_i} \rangle_* \biggr|_{\vec{\theta}_i \to \vec{\theta}_i + \vec{\theta}^\prime},
      \label{eq:meaningP} \\
  & & Q(\vec{\sigma}|\vec{\theta}^\prime) |_{\rm saddle}
      = \langle Q(\vec{\sigma}|\vec{\theta}^\prime) \rangle_*
      = \frac 1N \sum_{i=1}^N \biggl. \langle \delta_{\vec{\sigma},\vec{\sigma}_i} \rangle_* \biggr|_{k_i \to k_i - 1, \vec{\theta}_i \to \vec{\theta}_i + \vec{\theta}^\prime},
      \label{eq:meaningQ}
\end{eqnarray}
where the brackets $\langle \cdots \rangle_*$ denote evaluation of the argument for the microscopic process (\ref{eq:process}), i.e.
\begin{equation}
  \langle \cdots \rangle_*
  = \frac{\int \{\rmd P\rmd \hat{P}\rmd Q\rmd\hat{Q}\} \rme^{N \Psi[\{P,\hat{P},Q,\hat{Q}\}]} (\cdots)}
         {\int \{\rmd P\rmd\hat{P}\rmd Q\rmd\hat{Q}\} \rme^{N \Psi[\{P,\hat{P},Q,\hat{Q}\}]}}.
\end{equation}
Details on the derivation of (\ref{eq:meaningP}, \ref{eq:meaningQ}) can be found in \ref{app:physical_meaning}.
We may now conclude that $Q(\vec{\sigma}|\vec{\theta}^\prime)$ represents
the disorder-averaged probability of finding a single-spin trajectory $\vec{\sigma}$ in the system from which one site is removed randomly,
and in which the external field paths of spins $i$ previously connected to the removed site are modified to $\vec{\theta}_i+\vec{\theta}^\prime$. 
Similarly, $P(\vec{\sigma}|\vec{\theta}^\prime)$ represents
the disorder-averaged probability of finding a single-spin trajectory $\vec{\sigma}$ in the system 
in which the external field paths of a randomly drawn site are modified to $\vec{\theta}_i+\vec{\theta}^\prime$ (without removing sites). 
\par
\begin{figure}[t]
  \unitlength=0.22mm
  \begin{center}
    \begin{picture}(300,330)
      \thicklines
      \put(200,200){\circle{20}}\put(200,200){\circle{30}}\put(227,195){\large$\vec{\sigma}$}
      \put(200,290){\circle{30}}\put(225,285){\large$\vec{\sigma}_1$}
      \put(130,130){\circle{30}}\put(155,125){\large$\vec{\sigma}_2$}
      \put(270,130){\circle{30}}\put(297,125){\large$\vec{\sigma}_3$}
      \put(200,274){\vector(0,-1){58}}
      \put(142,142){\vector(1,1){46}}
      \put(258,142){\vector(-1,1){46}}
      \put(50,300){$\varepsilon Q( \vec{\sigma}_1| \frac{1}{c}J_1 \vec{\sigma} )$}
      \put(55,273){$+ (1\!-\!\varepsilon) P( \vec{\sigma}_1 | \vec{0})$}
      \put(116,195){$P( \vec{\sigma}|\vec{0})$}
      \put(-20,140){$\varepsilon Q( \vec{\sigma}_2| \frac{1}{c}J_2 \vec{\sigma} )$}
      \put(-15,113){$+ (1\!-\!\varepsilon) P( \vec{\sigma}_2 | \vec{0})$}
      \put(335,140){$\varepsilon Q( \vec{\sigma}_3| \frac{1}{c}J_3 \vec{\sigma} )$}
      \put(340,113){$+ (1\!-\!\varepsilon) P( \vec{\sigma}_3 | \vec{0})$}
    \end{picture}
    \vspace*{-20mm}
    \caption{
      Illustration of the interpretation of our closed-form order parameter equations, 
      showing how the measure for the spin paths $\vec{\sigma}$ of a central site with three incoming links 
      is related to the measures of the connected sites under modified conditions. 
      Each connected site has with probability $\varepsilon$  an incoming link from 
      the central site. For $N\to\infty$ our graphs are locally tree-like, 
      hence the three sites $i=1,2,3$ are mutually correlated only via the central site.
    }
    \label{fig:interpretation}
  \end{center}
\end{figure}
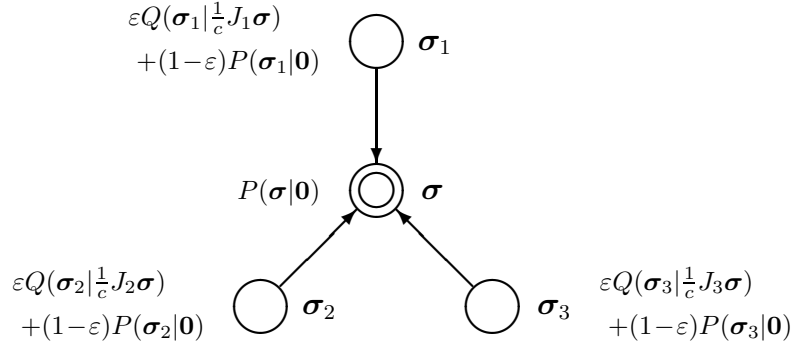

Our present general equations (\ref{eq:cP}, \ref{eq:cQ}) 
have a clearer interpretation than those obtained for strictly Poissonian graphs \cite{Hatchett2004}; 
in the latter graphs the distinction between (\ref{eq:meaningP}) and (\ref{eq:meaningQ}) is invisible 
as a consequence of the property  $\frac{k+1}{c} p(k+1)=p(k)$ of the Poissonian degree distribution. 
To calculate the probability of a single-site path $\vec{\sigma}$, a random number $k$ is firstly drawn from the degree distribution $p(k)$.
This parameter $k$ denotes the number of sites that contribute to the field at the central site.
All $k$ associated spin paths of the attached sites $\vec{\sigma}_1, \cdots, \vec{\sigma}_k$ are sampled 
from their respective distributions $\varepsilon  Q( \vec{\sigma}_\ell | \frac{1}{c} J_\ell \vec{\sigma} )+ (1\!-\!\varepsilon) P( \vec{\sigma}_\ell | \vec{0})$, 
which depend on the central site's path $\vec{\sigma}$. 
By definition, each site $\ell$ contributes to the field of the central site. 
Hence with probability $\varepsilon$ the central site will contribute $\frac{1}{c}J_\ell\vec{\sigma}$ 
to the field paths of site  $\ell$, hence the term $\varepsilon Q( \vec{\sigma}_\ell | \frac{1}{c} J_\ell \vec{\sigma} )$ in the path measure for $\ell$; 
with probability $1\!-\!\varepsilon$ the central site will not contribute to the fields of site  $\ell$, 
hence the term $(1\!-\!\varepsilon) P( \vec{\sigma}_\ell |\vec{0} )$ in the path measure for $\ell$.
This process makes it possible to take into account dynamically the effective retarded self-interaction induced by connection symmetry.
Figure \ref{fig:interpretation} shows a schematic illustration of this interpretation.

Equations (\ref{eq:meaningP}) and (\ref{eq:meaningQ}) show explicitly 
that the order parameters $P(\vec{\sigma}|\vec{\theta}^\prime)$ and $Q(\vec{\sigma}|\vec{\theta}^\prime)$ are normalized, 
i.e. that $\sum_{\vec{\sigma}} P(\vec{\sigma}|\vec{\theta}^\prime)=1$ and $\sum_{\vec{\sigma}} Q(\vec{\sigma}|\vec{\theta}^\prime)=1$. 
We have used this property to derive (\ref{eq:P^}) and (\ref{eq:Q^}).
In the special case of Poissonian graphs, i.e. graphs with $p(k)=e^{-c} c^k / k!$, the degree distribution obeys
\begin{equation}
  \frac{k+1}c p(k+1) = p(k)
\end{equation}
and hence one always has $P(\vec{\sigma}|\vec{\theta}^\prime)=Q(\vec{\sigma}|\vec{\theta}^\prime)$.
As a consequence our two closed-form order parameter equations reduce to a single closed equation:
\begin{eqnarray}
\hspace*{-10mm}
  P(\vec{\sigma}|\vec{\theta}^\prime)
  &=& \sum_{k=0}^\infty \frac{e^{-c} c^k}{k!} \biggl( \prod_{\ell=1}^k  \int\! \rmd J_\ell~ \tilde{p}(J_\ell) \sum_{\vec{\sigma}_\ell} \biggl[
           \varepsilon  P( \vec{\sigma}_\ell | \frac{J_\ell}{c} \vec{\sigma} )
      + (1\!-\!\varepsilon) P( \vec{\sigma}_\ell | \vec{0}            ) \biggr] \biggr) \nonumber \\
 \hspace*{-10mm} & & \times p_0 [\sigma(0)] \prod_{t=0}^{t_m\!-\!1}
      \frac{ \exp( \beta \sigma(t\!+\!1) [ \theta(t) \!+\! \theta^\prime(t) \!+\! \frac{1}{c} \sum_{\ell=1}^k\! J_\ell \sigma_\ell(t) ] ) }
           { 2 \cosh(\beta [ \theta(t) \!+\! \theta^\prime(t) \!+\! \frac{1}{c} \sum_{\ell=1}^k\! J_\ell \sigma_\ell(t) ] ) }.
\end{eqnarray}
and we thereby  recover the results of reference \cite{Hatchett2004}.

\section{Fully asymmetric connectivity}

\subsection{The reduced theory}
\par
We first consider the case of fully asymmetric connectivity, i.e.  $\varepsilon=0$.
Here the  situation is mathematically almost identical  to that case of the Poissonian graph \cite{Hatchett2004}.
Equation (\ref{eq:cP}) once more closes in terms of $P(\vec{\sigma}|\vec{0})$, and we find
\begin{eqnarray}
  P(\vec{\sigma}|\vec{0})
  &=& \sum_{k=0}^\infty p(k)
      \biggl( \prod_{\ell=1}^k  \int\! \rmd J_\ell~ \tilde{p}(J_\ell) \sum_{\vec{\sigma}_\ell} P( \vec{\sigma}_\ell | \vec{0}) \biggr) \nonumber \\
  & & \times p_0 [\sigma(0)] \prod_{t=0}^{t_m-1}
      \frac{ \exp( \beta \sigma(t\!+\!1) [ \theta(t) \!+\! \frac{1}{c} \sum_{\ell=1}^k J_\ell \sigma_\ell(t) ] ) }
           { 2 \cosh(\beta [ \theta(t) \!+\! \frac{1}{c} \sum_{\ell=1}^k J_\ell \sigma_\ell(t) ] ) }.
  \label{eq:P_e=0}
\end{eqnarray}
Summing both sides of (\ref{eq:P_e=0}) over $\vec{\sigma}$, except for the entry $\sigma(t+1)$,
leads us to single-time spin probabilities $P(\sigma(t+1)|\vec{0})$, which are marginal probabilities of the full path measure $P(\vec{\sigma}|\vec{0})$:
\begin{eqnarray}
  P(\sigma(t+1)|\vec{0})
  &=& \sum_{k=0}^\infty p(k)
      \biggl( \prod_{\ell=1}^k  \int\! \rmd J_\ell~ \tilde{p}(J_\ell) \sum_{\sigma_\ell(t)} P( \sigma_\ell(t) | \vec{0}) \biggr) \nonumber \\
  & & \times
      \frac{ \exp( \beta \sigma(t\!+\!1) [ \theta(t) \!+\! \frac{1}{c} \sum_{\ell=1}^k J_\ell \sigma_\ell(t) ] ) }
           { 2 \cosh(\beta [ \theta(t) \!+\! \frac{1}{c} \sum_{\ell=1}^k J_\ell \sigma_\ell(t) ] ) }.
  \label{eq:marginal_e=0}
\end{eqnarray}
It should be noted that the probabilities $P(\sigma(t+1)|\vec{0})$ depend only on $P(\sigma(t)|\vec{0})$, for any $t$.
This means that there is no effective retarded self-interaction,
i.e. there is no effective short loop in the graph with fully asymmetric connectivity.
Using the following two general identities
\begin{eqnarray}
 P(\sigma(t)|\vec{0}) &=& \frac{1}{2}[1 \!+\! \sigma(t) m(t)] \\
 P(\sigma(t),\sigma(t^\prime)|\vec{0}) &=& \frac{1}{4}[1\!+\!m(t)\sigma(t)\!+\!m(t^\prime)\sigma(t^\prime)
 \!+\!C(t,t^\prime)\sigma(t)\sigma(t^\prime)],
\end{eqnarray}
both the effective single spin magnetization $m(t) = \overline{\langle \sigma(t) \rangle}$ and the covariances $C(t,t^\prime) = \overline{\langle \sigma(t) \sigma(t^\prime) \rangle}$
can then be written as closed form iterative expressions:
\begin{eqnarray}
  \hspace*{-25mm}
  ~~~~~m(t+1)
      &=& \sum_{k=0}^\infty p(k)
      \biggl( \prod_{\ell=1}^k \sum_{\sigma_\ell}\! \frac{1}{2}[1 \!+\! \sigma_\ell m(t)] \biggr)
      \biggl\langle\! \tanh \biggl( \beta \biggl[ \theta(t) \!+\! \frac{1}{c} \sum_{\ell=1}^k J_\ell
      \sigma_\ell \biggr] \biggr) \!\biggr\rangle_{\!J_1,\cdots,J_k},\!\! \nonumber
      \\
  \hspace*{-25mm}
  & & \label{eq:m_eps=0}
  \\
  \hspace*{-25mm}
  C(t\!+\!1,t^\prime\!+\!1)
      &=& \sum_{k=0}^\infty p(k)
      \biggl( \prod_{\ell=1}^k \sum_{\sigma_\ell, \sigma_\ell^\prime} \frac 14[1\!+\!m(t)\sigma_\ell\!+\!m(t^\prime)\sigma_l^\prime
      \!+\!C(t,t^\prime) \sigma_\ell \sigma_\ell^\prime] \biggr)
      \nonumber \\
      \hspace*{-25mm}
  & & \hspace*{-0mm}\times
      \biggl\langle\!
      \tanh \biggl( \beta \biggl[ \theta(t) \!+\! \frac{1}{c} \sum_{\ell=1}^k J_\ell \sigma_\ell  \biggr] \biggr)
      \tanh \biggl( \beta \biggl[ \theta(t) \!+\! \frac{1}{c} \sum_{\ell=1}^k J_\ell \sigma_\ell^\prime \biggr] \biggr)
      \!\biggr\rangle_{\!J_1,\cdots,J_k}, \nonumber \\
   \hspace*{-25mm}
  & & \label{eq:C_eps=0}
\end{eqnarray}
from (\ref{eq:marginal_e=0}),
with $m(0)$ following from the initial conditions.
Figure \ref{fig:dynamics_eps=0} shows a comparison between theory and numerical simulations with respect to the time evolution of the magmetization $m(t)$ 
on a regular asymmetric sparse graph ($\varepsilon=0$) with the degree distribution $p(k)=\delta_{k,c}$. 
The theoretical results are in excellent agreement with the numerical simulations. 
\par
When taking the limit of $c\to\infty$, the internal fields $v_k(t) = \frac{1}{c} \sum_{l=1}^k J_\ell \sigma_\ell$ simplify to $v_k(t) \to \langle J \rangle_J m(t)$. In this limit we therefore have
\begin{eqnarray}
  & & m(t+1)=\tanh(\beta[\theta(t) + \langle J \rangle_J m(t)]), \\
  & & C(t,t^\prime)=m(t)m(t^\prime).
\end{eqnarray}
In the absence of external fields, i.e. for $\theta(t)=0$, a P$\to$F transition occurs at $\beta=\beta_c$, which is given by $\beta_c \langle J \rangle_J =1$.
This situation is identical to that described in \cite{Hatchett2004}.

\subsection{Time evolution and phase diagrams}

We here specialize further and consider the physics described by the $\varepsilon=0$ equations
 (\ref{eq:m_eps=0}, \ref{eq:C_eps=0}) for the case of having binary random bonds:
\begin{equation}
  \tilde{p}(J^\prime) = \frac{1}{2} (1\!+\!\eta) \delta(J^\prime\!-J) + \frac{1}{2}(1\!-\!\eta) \delta(J^\prime+J),
  \label{eq:random_bond}
\end{equation}
with $\eta\in[-1,1]$.
For such bond statistics the iterative equation (\ref{eq:m_eps=0}) for the magnetization reduces to
\begin{eqnarray}
  m(t+1)
  &=& \sum_{k=0}^\infty p(k)
      \sum_{r=0}^k \Big(\!\!
      \begin{array}{c}
        k \\
        r
      \end{array}\!\!\Big)
      \Big(\! \frac{1\!+\!\eta m(t)}2 \!\Big)^{\!r}
      \Big(\! \frac{1\!-\!\eta m(t)}2 \!\Big)^{\!k-r} \nonumber \\
  & & \times \tanh \Big( \beta \Big[ \theta(t) + \frac{J}{c} (2r\!-\!k) \Big] \Big)
\end{eqnarray}
For zero external fields, i.e. $\theta(t)=0$ for all $t$, and
upon assuming that a stationary state exists, the stationary state magnetizations are given as the solutions of the fixed-point equation
$m=F(m)$, with
\begin{eqnarray}
 \hspace*{-5mm} F(m)
  &=& \sum_{k=0}^\infty p(k)
      \sum_{r=0}^k
      \Big(\!\!\begin{array}{c} k \\ r\end{array}\!\!\Big)
      \Big(\! \frac{1\!+\!\eta m}2 \!\Big)^{\!r}
      \Big(\! \frac{1\!-\!\eta m}2 \!\Big)^{\!k-r} \! \tanh \Big(\frac{\beta J(2r\!-\!k)}{c}\Big)
\end{eqnarray}
The map $F(m)$ is anti-symmetric, and hence always has the trivial fixed-point $m=0$.
This fixed-point is unique for small $\eta$, whereas for larger $\eta$ nontrivial fixed-points bifurcate (provided the inverse temperature $\beta$ is sufficiently large), marking a transition P$\to$F from a paramagnetic to a ferro-megnetic state.
To determine whether this transition is continuous, and if so find the critical value $\eta_c$,
we can expand $F(m)$ in powers of $m$, giving
\begin{eqnarray}
\hspace*{-20mm}
  F(m)
  &=& \eta m
      \sum_{k=0}^\infty p(k)
      \sum_{r=0}^k \Big(\!\!\begin{array}{c} k \\ r\end{array}\!\!\Big)
      \frac{2r\!-\!k}{2^k} \tanh \Big[ \frac{\beta J (2r\!-\!k)}{c} \Big] \nonumber
      \\
   \hspace*{-20mm}
  & & \hspace*{-3mm}+ \frac 16 (\eta m)^3
      \!\sum_{k=0}^\infty p(k)
      \sum_{r=0}^k
      \Big(\!\!\begin{array}{c} k \\ r\end{array}\!\!\Big) \frac{(2r\!-\!k)(4r^2\!\!-\!4kr\!+\!k^2\!\!-\!3k\!+\!2)}{2^k}
      \tanh \Big[ \frac{\beta J (2r\!-\!k)}{c} \Big] \nonumber \\
 \hspace*{-20mm} & & + \mathcal{O}(m^5)
\end{eqnarray}
The cubic term can be confirmed to be non-positive.
Therefore there is no evidence for a discontinuous transition, and the critical value $\eta_c$ can be obtained as
\begin{eqnarray}
  \eta_c
  \sum_{k=0}^\infty p(k)
  \sum_{r=0}^k \Big(\!\!\begin{array}{c} k \\ r\end{array}\!\!\Big)
  \frac{2r\!-\!k}{2^k} \tanh \Big[ \frac{\beta}{c} J (2r\!-\!k) \Big] =1.
\end{eqnarray}
In a similar way we can inspect the existence and location of a spin-glass phase.
Upon putting $m(t)=0$ and $\theta(t)=0$ in (\ref{eq:C_eps=0}),
one finds that the time-translation invariant covariance $q=\lim_{\tau\to\infty}\lim_{t\to\infty}C(t+\tau,t)$ are given as the solutions of $q=G(q)$, with
\begin{eqnarray}
\hspace*{-8mm}
  G(q)
  &=& \sum_{k=0}^\infty p(k)
      \biggl( \prod_{\ell=1}^k \sum_{\sigma_\ell,\sigma_\ell^\prime} \frac{1}{4}[1\!+\!\sigma_\ell \sigma_\ell^\prime q]\biggr)
      \tanh \biggl[ \frac{\beta J}{c} \sum_{\ell=1}^k \sigma_\ell  \biggr]
      \tanh \biggl[ \frac{\beta J}{c} \sum_{\ell=1}^k \sigma_\ell^\prime \biggr]. \nonumber \\
      \hspace*{-8mm}&&
\end{eqnarray}
It can be confirmed that $G(0)=0$, $G(q) \le 1$ and $\rmd^2G(q)/\rmd q^2>0$ for $q \in [0,1]$.
Hence $G(q)<q$ for $q \in (0,1]$.
Since $q=G(q)$ has no non-trivial solutions, no spin-glass phase exists for $\varepsilon=0$. 
\par
\begin{figure}[t]
  \begin{center}
    \includegraphics[width=250\unitlength,keepaspectratio]{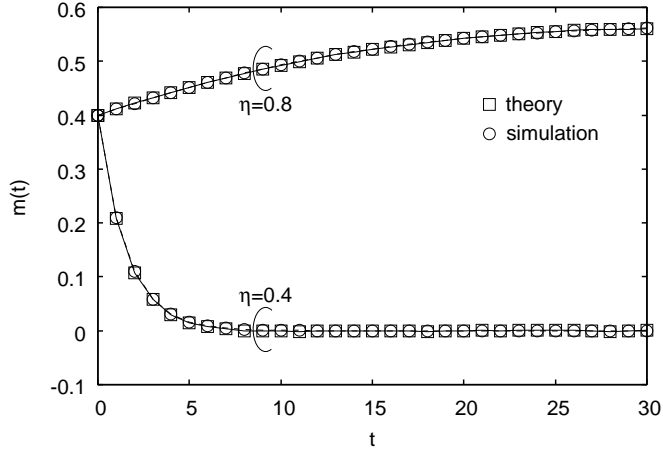}
    \caption{
        Comparison between theory and numerical simulations with respect to the time evolution of the magmetization $m(t)$ 
        on a asymmetric $(\varepsilon=0)$ finitely connected random graph with a degree distribution $p(k)=\delta_{k,c}$ and zero external fields $\theta(t)=0$. 
        The initial magnetization is $m(0)=0.4$. 
        Squares: theoretical results for $c=3$, $\beta=3$ and $\eta\in\{0.4, 0.8\}$.
        Circles: simulation resuls for $N=10^5$ spins (averaged over 10 runs).
    }
    \label{fig:dynamics_eps=0}
  \end{center}
\end{figure}

\begin{figure}[t]
  \begin{center}
    \includegraphics[width=250\unitlength,keepaspectratio]{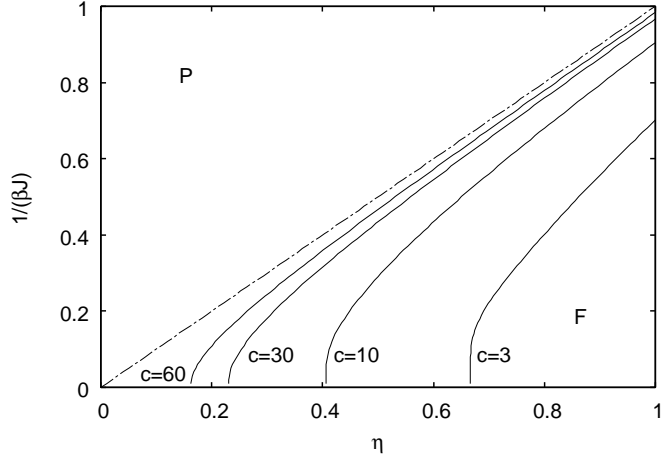}
    \caption{
        Phase diagram in the $(\eta,1/(\beta J))$ plane of the $\pm J$ random bond model on a asymmetric ($\varepsilon=0$) sparse graph 
        with the degree distribution $p(k)=\delta_{k,c}$ and and zero external fields $\theta(t)=0$.
        Solid lines: P $\to$ F transition lines for $c\in\{3,10,30,60\}$.
        Dashed line: P $\to$ F transition line at $1/(\beta J)=\eta$ corresponding to $c=\infty$.
    }
    \label{fig:phase}
  \end{center}
\end{figure}

 
Figure \ref{fig:dynamics_eps=0} shows a comparison between theory and numerical simulations with respect to the time evolution of the magmetization $m(t)$ 
on a regular asymmetric sparse graph ($\varepsilon=0$) with the degree distribution $p(k)=\delta_{k,c}$ and zero external fields $\theta(t)=0$. 
The theoretical results are in excellent agreement with the numerical simulations.  
Figure \ref{fig:phase} shows the resulting phase diagram in the $(\eta,1/\beta J)$ plane, 
for the example of a regular sparse graph with the degree distribution $p(k)=\delta_{k,c}$ and zero external fields $\theta(t)=0$.
For nonzero temperatures, the $c=2$ regular sparse graph has only a paramagnetic phase (with a zero temperature P$\to$F transition only for $\eta=1$).

\section{Arbitrary connectivity symmetry}

\subsection{Numerical solution for short times}

\begin{figure}[t]
  \begin{center}
    \includegraphics[width=250\unitlength,keepaspectratio]{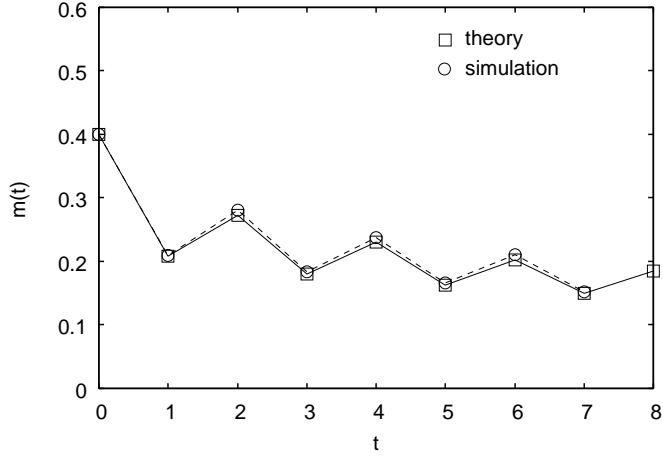}
    \caption{
        Comparison between theory and numerical simulations with respect to the time evolution of the magmetization $m(t)$ 
        on a symmetric $(\varepsilon=1)$ finitely connected random graph with a degree distribution $p(k)=\delta_{k,c}$. 
        The initial magnetization is $m(0)=0.4$. 
        Squares: theoretical results for $c=3$, $\beta=3$ and $\eta=0.4$. 
        Circles: simulation resuls for $N=10^5$ spins (averaged over 10 runs). 
    }
    \label{fig:dynamics_eps=1}
  \end{center}
\end{figure}

Equations (\ref{eq:cP}) and (\ref{eq:cQ}) are closed and exact, but highly nontrivial, 
and it is not obvious that they can be simplified further.
However, if the bond distribution is of the form (\ref{eq:random_bond}), 
the space on which our equations are defined at least becomes finite dimensional. 
In that particular case we can achieve closure for the following reduced order parameters:
\begin{eqnarray}
  & & {\sf P}(\vec{\sigma})=P(\vec{\sigma}| \vec{0} ),~~~~~~~~ {\sf Q}(\vec{\sigma}|\vec{\sigma}^\prime)=Q(\vec{\sigma}| \frac{J}{c} \vec{\sigma}^\prime ),
\end{eqnarray}
since from equations (\ref{eq:cP}, \ref{eq:cQ}) one can extract
\begin{eqnarray}
\hspace*{-15mm}
  {\sf P}(\vec{\sigma})
  &=& \sum_{k=0}^\infty p(k) \biggl( \prod_{\ell=1}^k \!\sum_{\tau_\ell=\pm 1}\! \frac{1}{2}[1\!+\!\eta \tau_\ell] \sum_{\vec{\sigma}_\ell}[
           \varepsilon  {\sf Q}( \vec{\sigma}_\ell | \tau_\ell \vec{\sigma} )
     \! +\! (1\!-\!\varepsilon) {\sf P}( \vec{\sigma}_\ell ) ] \biggr) \nonumber
     \\
  \hspace*{-15mm}
  & & \times p_0 [\sigma(0)] \prod_{t=0}^{t_m-1}
      \frac{ \exp( \beta \sigma(t\!+\!1) [ \theta(t) \!+\! \frac{J}{c} [ \sigma^\prime(t) \!+\! \sum_{\ell=1}^k \tau_\ell \sigma_\ell(t) ] ) }
           { 2 \cosh(\beta [ \theta(t) \!+\! \frac{J}{c} [ \sigma^\prime(t) \!+\! \sum_{\ell=1}^k \tau_\ell \sigma_\ell(t) ] ) },
  \label{eq:cP2}
  \\
  \hspace*{-15mm}  {\sf Q}(\vec{\sigma}|\vec{\sigma}')
  &=& \sum_{k=0}^\infty \frac{k\!+\!1}{c} p(k\!+\!1) \biggl( \prod_{\ell=1}^k \!\sum_{\tau_\ell=\pm 1}\!
   \frac{1}{2}[1\!+\!\eta \tau_\ell] \sum_{\vec{\sigma}_\ell}[
           \varepsilon  {\sf Q}( \vec{\sigma}_\ell | \tau_\ell \vec{\sigma} )
      \!+\! (1\!-\!\varepsilon) {\sf P}( \vec{\sigma}_\ell ) ] \biggr) \nonumber
      \hspace*{-15mm}
      \\
      \hspace*{-15mm}
  & & \times p_0 [\sigma(0)] \prod_{t=0}^{t_m-1}
      \frac{ \exp( \beta \sigma(t\!+\!1) [ \theta(t) \!+\! \frac{J}{c} [ \sigma^\prime(t) \!+\! \sum_{\ell=1}^k \tau_\ell \sigma_\ell(t) ] ) }
           { 2 \cosh(\beta [ \theta(t) \!+\! \frac{J}{c} [ \sigma^\prime(t) \!+\! \sum_{\ell=1}^k \tau_\ell \sigma_\ell(t) ] ) },
  \label{eq:cQ2}
\end{eqnarray}
with $\vec{\sigma}^\prime \in \{-1,1\}^{t_m}$.
Equations (\ref{eq:cP2}) and (\ref{eq:cQ2}) can be solved numerically by iteration.
The order parameter fucntion ${\sf P}(\vec{\sigma})$ is the disorder-averaged probability of finding a single-spin trajectory 
$\vec{\sigma}=(\sigma(0), \cdots, \sigma(t), \cdots, \sigma(t_m))$ with zero external fields. 
Therefore the magnetization is given by
\begin{equation}
  m(t) = \sum_{\vec{\sigma}} \sigma(t) {\sf P}(\vec{\sigma}).
\end{equation}
Figure \ref{fig:dynamics_eps=1} shows a comparison between theory and numerical simulations with respect to the time evolution of the magnetization $m(t)$
on a regular symmetric sparse graph, i.e. $\varepsilon=1$ and the degree distribution is $p(k)=\delta_{k,c}$.
The theoretical results are in good agreement with the numerical simulations.
In the case $\varepsilon=1$, the numerical analysis in fact simplifies slightly:
since (\ref{eq:cQ2}) no longer involves ${\sf P}(\vec{\sigma})$, one can first solve (\ref{eq:cQ2})  for ${\sf Q}(\vec{\sigma}|\vec{\sigma}^\prime)$,
and then substitute the solution into (\ref{eq:cP2}) to obtain ${\sf P}(\vec{\sigma})$.

\subsection{Approximate stationary solutions for fully symmetric connectivity}
\par
We next consider approximate stationary solutions of our macroscopic equations.
Detailed balance holds only for $\varepsilon=1$.
Since we here discuss dynamics with parallel updates, the equilibrium state for $\varepsilon=1$ does not have a Boltzmann form but involves Peretto's pseudo-Hamiltonian \cite{Peretto1984}.
In the equilibrium state initial conditions are required to be irrelevant, therefore we may shift the initial and final times to $-\infty$ and $\infty$, respectively. 
We also choose zero external fields, $\theta(t)=0$.
To find approximate stationary solutions, we propose the following ansatz:
\begin{eqnarray}
  P(\vec{\sigma}|\vec{\theta}^\prime) &=& \int\! \rmd h~ \mathcal{P}(h) \prod_t \frac{e^{\beta \sigma(t+1)[h+\theta'(t)]}}{2\cosh(\beta [h+\theta'(t)])}, \label{eq:ansatzP} \\
  Q(\vec{\sigma}|\vec{\theta}^\prime) &=& \int\! \rmd h~ \mathcal{Q}(h) \prod_t \frac{e^{\beta \sigma(t+1)[h+\theta'(t)]}}{2\cosh(\beta [h+\theta'(t)])}, \label{eq:ansatzQ}
\end{eqnarray}
where $\mathcal{P}(h)$ denotes an effective local field distribution, and $\mathcal{Q}(h)$ denotes an effective cavity field distribution \cite{Mezard1987b}.
Substituting this ansatz into the right-hand sides of (\ref{eq:cP}) and (\ref{eq:cQ}) results for large $c$ in
(see \ref{app:approx_stationary_state} for details):
\begin{eqnarray}
  P(\vec{\sigma}|\vec{\theta}^\prime)
  &=& \int\! \rmd h~
      \biggl\{
        \sum_{k=0}^\infty p(k)
        \biggl( \prod_{\ell=1}^k \int\! \rmd h_\ell\rmd J_\ell~ \tilde{p}(J_\ell) \mathcal{Q}(h_\ell) \biggr)
         \nonumber \\
  & &   \times \delta \biggl( h - \frac 1\beta \sum_{\ell=1}^k \tanh^{-1} [ (\tanh \beta h_\ell) (\tanh \frac{\beta J}{c}) ] \biggr)
      \biggr\} \nonumber
      \\
  & & \times      \prod_t \frac{e^{\beta \sigma(t+1)[h+\theta^\prime(t)]}}{2\cosh(\beta [h\!+\!\theta^\prime(t)])}
      \label{eq:ss_P} \\
  Q(\vec{\sigma}|\vec{\theta}^\prime)
  &=& \int\! \rmd h~
      \biggl\{
        \sum_{k=0}^\infty \frac{k\!+\!1}{c} p(k\!+\!1)
        \biggl( \prod_{\ell=1}^k \int\! \rmd h_\ell \rmd J_\ell~ \tilde{p}(J_\ell) \mathcal{Q}(h_\ell) \biggr)
        \nonumber \\
  & &   \times \delta \biggl( h - \frac 1\beta \sum_{\ell=1}^k \tanh^{-1} [ (\tanh \beta h_\ell) (\tanh \frac{\beta J}{c}) ] \biggr)
      \biggr\} \nonumber \\
  && \times
     \prod_t \frac{e^{\beta \sigma(t+1)[h+\theta'(t)]}}{2\cosh(\beta [h\!+\!\theta^\prime(t)])}
  \label{eq:ss_Q}
\end{eqnarray}
Comparing (\ref{eq:ss_P}, \ref{eq:ansatzP}) to (\ref{eq:ss_Q}, \ref{eq:ansatzQ}) then implies the
following relationships for the effective field distributions $\mathcal{P}(h)$ and $\mathcal{Q}(h)$:
\begin{eqnarray}
  \mathcal{P}(h)
  &=& \sum_{k=0}^\infty p(k)
        \biggl( \prod_{\ell=1}^k \int\! \rmd h_\ell \rmd J_\ell~ \tilde{p}(J_\ell) \mathcal{Q}(h_\ell) \biggr) \nonumber \\
  & &   \times \delta \biggl( h - \frac 1\beta \sum_{\ell=1}^k \tanh^{-1} [ (\tanh \beta h_\ell) (\tanh \frac{\beta J}{c}) ] \biggr), \label{eq:lfd_P} \\
  \mathcal{Q}(h)
  &=& \sum_{k=0}^\infty \frac{k\!+\!1}{c} p(k\!+\!1)
        \biggl( \prod_{\ell=1}^k \int\! \rmd h_\ell \rmd J_\ell ~\tilde{p}(J_\ell) \mathcal{Q}(h_\ell) \biggr) \nonumber \\
  & &   \times \delta \biggl( h - \frac 1\beta \sum_{\ell=1}^k \tanh^{-1} [ (\tanh \beta h_\ell) (\tanh \frac{\beta J}{c}) ] \biggr). \label{eq:lfd_Q}
\end{eqnarray}
Equation (\ref{eq:lfd_Q}) has a closed form with respect to $\mathcal{Q}(h)$; its solution
is substituted  into (\ref{eq:lfd_P}) to generate $\mathcal{P}(h)$.
In the special case of Poissonian graphs, i.e. $p(k)=e^{-c} c^k / k!$, we observe that
$\mathcal{P}(h)=\mathcal{Q}(h)$, and we therefore have just one closed equation:
\begin{eqnarray}
  \mathcal{P}(h)
  &=& \sum_{k=0} \frac{e^{-c} c^k}{k!}
        \biggl( \prod_{\ell=1}^k \int\! \rmd h_\ell \rmd J_\ell~ \tilde{p}(J_\ell) \mathcal{P}(h_\ell) \biggr) \nonumber \\
  & &   \times \delta \biggl( h - \frac 1\beta \sum_{\ell=1}^k \tanh^{-1} [ (\tanh \beta h_\ell) (\tanh \frac{\beta J}{c}) ] \biggr).
\end{eqnarray}
We thereby  recover for Poissonnian graphs the result of \cite{Hatchett2004}, 
which is, in turn, identical to the replica symmetric equilibrium solution of the sequential dynamics version of the model addressed here \cite{Mezard1987c}. 
In the reference \cite{Castillo2004a}, it has been shown that one expects the replica symmetric equilibrium solutions of sequential and parallel dyamics to be identical. 

\section{Discussion}
\par
In this paper, we have applied the generating functional analysis technique
to the dynamics of Ising spin models on finitely connected random graph with arbitrary degree distributions,
following in the footsteps of reference \cite{Hatchett2004} (which was limited to Poissonnian degree distributions).
We have first derived general exact equations to represents dynamical properties of the system in the infinite size limit.
The introduction of arbitrary degree distributions was found to give us very clear and intuitive interpretations of the macroscopic dynamics of finitely connected Ising systems, 
in terms of macroscopic path probability distributions involving the evolution of both the actual local field and the cavity field. 
We have already applied our theory to error correcting codes \cite{Mimura2009b}. 
We next aim to apply the theory to other problems in the field of information theory, 
as well as generalize it to include more complicated local field definitions which will enable it to be used for the analysis of the dynamics of gene regulation systems. 
Especially in the latter systems, where the graphs concerned are finitely connected and directed but where the degree distributions are known to be far from Poissonnian, 
the ability to study dynamics macroscopically for arbitrary degree distributions is vital.

\ack
This work was partially supported by a Grant-in-Aid for Encouragement of Young Scientists (B) No. 18700230
from the Ministry of Education, Culture, Sports, Science and Technology of Japan.

\section*{References}

\appendix

\section{Calculation of the disorder average \label{app:disorder_average}} 
\par
We here calculate the disorder average. 
The term containing the disorder, which is the left-hand side of (\ref{eq:DisorderTerm}), becomes 

\begin{eqnarray}
  \hspace*{-15mm}
  & & \hspace*{-10mm} \overline{ \exp \biggl[ - \frac{\rmi}{c} \sum_{i=1}^N \sum_{t=0}^{t_m-1} \hat{h}_i(t) \sum_{j \ne i}^N c_{ij} J_{ij} \sigma_j(t) \biggr] } \nonumber \\
  \hspace*{-20mm} &=& \frac1{Z_c}
      \sum_{\vec{c}} \biggl( \prod_{i=1}^N \prod_{j>i}^N \hat{p}(c_{ij}) \hat{p}(c_{ji}|c_{ij}) \int \rmd J_{ij} \tilde{p}(J_{ij}) \nonumber \\
  \hspace*{-20mm} & & \times \exp \biggl[ - \frac{\rmi}c \sum_{t=0}^{t_m-1} [\hat{h}_i(t) c_{ij} J_{ij} \sigma_j(t) + \hat{h}_j(t) c_{ji} J_{ji} \sigma_i(t) ] \biggr] \biggr) \nonumber \\
  \hspace*{-20mm} & & \times \biggl( \prod_{i=1}^N \int_0^{2\pi} \frac{d \omega_i}{2\pi} \exp \biggl[ \rmi \omega_i \biggl( k_i - \sum_{j=1}^N c_{ij} \biggr) \biggr] \biggr) \nonumber \\
  \hspace*{-20mm} &=& \frac1{Z_c}
      \biggl( \prod_{i=1}^N \int_0^{2\pi} \frac{d \omega_i}{2\pi} e^{\rmi \omega_i k_i} \biggr)
      \biggl( \prod_{i=1}^N \prod_{j>i}^N \biggl\langle \sum_{c_{ij},c_{ji}} \hat{p}(c_{ij}) \hat{p}(c_{ji}|c_{ij}) \nonumber \\
  \hspace*{-20mm} & & \times \exp \biggl[ - \rmi \frac Jc \sum_{t=0}^{t_m-1} [\hat{h}_i(t) c_{ij} \sigma_j(t) + \hat{h}_j(t) c_{ji} \sigma_i(t) ] 
                      - \rmi (\omega_i c_{ij} + \omega_j c_{ji}) \biggr] \biggr\rangle_J \biggr). 
\end{eqnarray}
We then have (\ref{eq:DisorderTerm}).

\section{Derivation of the physical meaning of our order parameters \label{app:physical_meaning}}
\par
The physical meaning of the order parameter functions can be inferred by evaluating (\ref{eq:meaningP}) and (\ref{eq:meaningQ}).
We first present the outline of the derivation of the physical meaning of $Q(\vec{\sigma}|\vec{\theta}^\prime)$.
Equation (\ref{eq:meaningQ}) becomes
\begin{eqnarray}
  \hspace*{-20mm}
  Q(\vec{\sigma}|\vec{\theta}^\prime) |_{\rm saddle} 
  &=& \langle Q(\vec{\sigma}|\vec{\theta}^\prime) \rangle_* \nonumber \\
      \hspace*{-20mm}  
  & & \hspace*{-20mm} 
      = \frac
      {\displaystyle \int\! \{\rmd P\rmd\hat{P}\rmd Q\rmd\hat{Q}\} \rme^{N \Psi[\{P,\hat{P},Q,\hat{Q}\}]}
       \Big( \int\! \rmd\hat{\vec{h}}~ \rme^{ -\rmi \vec{\theta}^\prime \cdot \hat{\vec{h}} } \frac 1N \sum_{i=1}^N \delta_{\vec{\sigma},\vec{\sigma}_i} \delta(\hat{\vec{h}}-\hat{\vec{h}}_i) \rme^{-\rmi \omega_i} \Big)}
      {\displaystyle \int\! \{\rmd P\rmd \hat{P}\rmd Qd\hat{Q}\} \rme^{N \Psi[\{P,\hat{P},Q,\hat{Q}\}]}} 
      \nonumber \\
      \hspace*{-20mm}
  & & \hspace*{-20mm} 
      =\biggl[ \int\! \{\rmd P\rmd\hat{P}\rmd Q\rmd\hat{Q}\} \rme^{N \Psi[\{P,\hat{P},Q,\hat{Q}\}]} \biggr]^{-1} 
      \biggl[
        \frac 1N \sum_{i=1}^N
        \int\! \{\rmd P\rmd\hat{P}\rmd Q\rmd\hat{Q}\} \rme^{N \Omega[\{P,\hat{P},Q,\hat{Q}\}]} 
        \nonumber\\
        \hspace*{-20mm}&&\times
        \biggl\{ \prod_{j=1}^N \biggl( \sum_{\vec{\sigma}_j} \int\! \rmd\vec{h}_j \rmd\hat{\vec{h}}_j \int\! \rmd\omega_j \rme^{\rmi \omega_j k_j}
        \rho_j(\vec{\sigma}_j,\vec{h}_j,\hat{\vec{h}}_j,\vec{\theta}_i,\omega_j) \biggr) \biggr\} \nonumber \\
  & &   \qquad \times
        \int\! \rmd\hat{\vec{h}} ~\rme^{ -\rmi \vec{\theta}^\prime \cdot \hat{\vec{h}} } \delta_{\vec{\sigma},\vec{\sigma}_i} \delta(\hat{\vec{h}}-\hat{\vec{h}}_i) \rme^{-\rmi \omega_i}
      \biggr],
  \label{eq:meaningQ_1}
\end{eqnarray}
where $\rho_i(\vec{\sigma}_i,\vec{h}_i,\hat{\vec{h}}_i,\vec{\theta}_i,\omega_i)$ and $\Omega(\{P,\hat{P},Q,\hat{Q}\})$ are defined as follows:
\begin{eqnarray}
  \rho_i(\vec{\sigma}_i,\vec{h}_i,\hat{\vec{h}}_i,\vec{\theta}_i,\omega_i)
  &=& \frac 1{(2\pi)^{t_m+1}}
      \rme^{-\rmi\hat{P}(\vec{\sigma}_i,\hat{\vec{h}}_i)}
      \rme^{-\rmi\hat{Q}(\vec{\sigma}_i,\hat{\vec{h}}_i)\rme^{-\rmi\omega_i}}  \\
  & & \times p_0[\sigma_i(0)] \prod_{t=0}^{t_m-1} \frac {\rme^{\rmi\hat{h}_i(t)[h_i(t)-\theta_i(t)]+\beta \sigma_i(t+1) h_i(t)}}{2 \cosh [\beta h_i(t)]},\nonumber
\end{eqnarray}\vspace*{-2mm}
\begin{eqnarray}
  \Omega[\{P,\hat{P},Q,\hat{Q}\}]
  &=& \frac c2(\varepsilon\!-\!2)
      + \frac c2 \sum_{\vec{\sigma}} \sum_{\vec{\sigma}^\prime} \int\! \rmd\hat{\vec{h}}\rmd\hat{\vec{h}^\prime}~ A(\vec{\sigma},\hat{\vec{h}};\vec{\sigma}^\prime,\hat{\vec{h}}^\prime)
      \nonumber \\
  & & \hspace*{-10mm}
      + \rmi \sum_{\vec{\sigma}} \int\! \rmd\hat{\vec{h}}\Big[ \hat{P}(\vec{\sigma},\hat{\vec{h}})P(\vec{\sigma},\hat{\vec{h}})  
      +\hat{Q}(\vec{\sigma},\hat{\vec{h}}) Q(\vec{\sigma},\hat{\vec{h}}) \Big] . 
\end{eqnarray}
The term in (\ref{eq:meaningQ_1}) that involves $\vec{\omega}$-integrations can be evaluated as follows:
\begin{eqnarray}
  & & \hspace*{-25mm} \biggl\{ \prod_{j=1}^N \biggl( \sum_{\vec{\sigma}_j} \int\! \rmd\vec{h}_j \rmd\hat{\vec{h}}_j \rmd\omega_j~ \rme^{\rmi \omega_j k_j}
      \rho_j(\vec{\sigma}_j,\vec{h}_j,\hat{\vec{h}}_j,\vec{\theta}_i,\omega_j) \biggr) \biggr\} 
      \int\! \rmd\hat{\vec{h}}~ \rme^{ -\rmi \vec{\theta}^\prime\cdot \hat{\vec{h}} } \delta_{\vec{\sigma},\vec{\sigma}_i} \delta(\hat{\vec{h}}-\hat{\vec{h}}_i) \rme^{-\rmi \omega_i} \nonumber \\
  & & \hspace*{-10mm} = \biggl\{ \prod_{j \ne i}^N \biggl( \sum_{\vec{\sigma}_j} \int\! \rmd\vec{h}_j \rmd\hat{\vec{h}}_j \rmd\omega_j~ \rme^{\rmi \omega_j k_j}
      \rho_j(\vec{\sigma}_j,\vec{h}_j,\hat{\vec{h}}_j,\vec{\theta}_i,\omega_j) \biggr) \biggr\} \nonumber \\
  & & \hspace*{-10mm} ~~~ \times 
      \sum_{\vec{\sigma}_i}
      \int\! \rmd\vec{h}_i\rmd\hat{\vec{h}}\rmd\omega_i
      ~\rme^{\rmi \omega_i (k_i-1)}
      \rho_i(\vec{\sigma}_i,\vec{h}_i,\hat{\vec{h}},\vec{\theta}_i+\vec{\theta}^\prime\!,\omega_i)
      \delta_{\vec{\sigma},\vec{\sigma}_i} \nonumber\\
  & & \hspace*{-10mm} = \biggl.
      \biggl\{
        \prod_{j=1}^N \biggl( \sum_{\vec{\sigma}_j} \int\! \rmd\vec{h}_j \rmd\hat{\vec{h}}_j \rmd\omega_j~ \rme^{\rmi \omega_j k_j}
        \rho_j(\vec{\sigma}_j,\vec{h}_j,\hat{\vec{h}}_j,\vec{\theta}_i,\omega_j) \biggr)
      \biggr\}
      \biggr|_{k_i \to k_i - 1, \vec{\theta}_i \to \vec{\theta}_i + \vec{\theta}'}. \nonumber \\
  \label{eq:meaningQ_2}
\end{eqnarray}
Substituting (\ref{eq:meaningQ_2}) into (\ref{eq:meaningQ_1}) then gives us (\ref{eq:meaningQ}).
Equation (\ref{eq:meaningP}) can be obtained in a similar way, since the difference is just the absence of the factor $\rme^{-\rmi \omega_i}$.

\section{Derivation of approximate stationary solutions \label{app:approx_stationary_state}}
\par
Equations (\ref{eq:ss_P}) and (\ref{eq:ss_Q}) are obtained following the reasoning in \cite{Hatchett2004}. 
We first derive (\ref{eq:ss_P}), using two specific identities. The first applies to $\sigma \in \{-1,1\}$: 
\begin{equation}
  \cosh[\beta(a+b\sigma)] = A e^{\beta B \sigma},
  \label{eq:id1}
\end{equation}
with
\begin{eqnarray}
  \hspace*{-10mm} A=\sqrt{\cosh[\beta(a\!+\!b)] \cosh[\beta(a\!-\!b)]}, ~~~~~~
  B=\frac 1{2\beta} \ln \frac{\cosh[\beta(a\!+\!b)]}{\cosh[\beta(a\!-\!b)]},
\end{eqnarray}
The second identity holds for $\sigma \in \{-1,0,1\}$ and any function $f(\sigma)$:
\begin{equation}
  f(\sigma)=Ce^{\beta D \sigma + \beta E \sigma^2}, 
  \label{eq:id2}
\end{equation}
with 
\begin{eqnarray}
  C=f(0), ~~~~~~
  D=\frac 1{2\beta} \ln \frac{f(1)}{f(-1)}, ~~~~~~
  E=\frac 1{2\beta} \ln \frac{f(1)f(-1)}{f(0)^2}.
\end{eqnarray}
We abbreviate $\prod_t [\frac 12 \sum_{\sigma(t)=\pm 1} f(\vec{\sigma})] = \langle f(\sigma) \rangle_{\vec{\sigma}}$ and
substitute the ansatz (\ref{eq:ansatzQ}) into the right-hand side of (\ref{eq:cP}) with $\varepsilon=1$, which gives 
\begin{eqnarray}
  \hspace*{-15mm}
  P(\vec{\sigma}|\vec{\theta}^\prime)
  &=& \sum_{k=0}^\infty p(k)
      \biggl\langle \biggl( \prod_{\ell=1}^k  \int\! \rmd J_\ell~ \tilde{p}(J_\ell)
      \int\! \rmd h~ \mathcal{Q}(h) \prod_t \frac{\rme^{\beta \sigma(t+1)[h+\theta^\prime(t)]}}{\cosh(\beta [h+\theta^\prime(t)])}
      \biggr) \nonumber \\
  \hspace*{-15mm}
  & & \times \prod_t
      \frac{ \exp( \beta \sigma(t\!+\!1) [ \theta^\prime(t) + \frac{1}{c}\sum_{\ell=1}^k J_\ell \sigma_\ell(t) ] ) }
           { 2 \cosh(\beta [ \theta^\prime(t) + \frac{1}{c} \sum_{\ell=1}^k J_\ell \sigma_\ell(t) ] ) } \biggr\rangle_{\vec{\sigma}} \nonumber 
           \\
  \hspace*{-15mm}           
  &=& \sum_{k=0}^\infty p(k)
      \biggl( \prod_{\ell=1}^k  \int\! dJ_\ell~ \tilde{p}(J_\ell) \int\! \rmd h~ \mathcal{Q}(h) \biggr) 
       \biggl\{ \prod_t \frac{\rme^{\beta \sigma(t+1)[\theta^\prime(t)-\sum_{\ell=1}^k B_\ell]}}{\prod_{\ell=1}^k A_\ell} \biggr\} \nonumber \\
  \hspace*{-15mm}  
  & & \times \prod_t \biggl\langle
      \frac{ \exp( \beta \sum_{\ell=1}^k \sigma_\ell [ h_\ell+\frac{1}{c}J_\ell \{ \sigma(t\!-\!1)+\sigma(t\!+\!1) \} ] ) }
           { 2 \cosh(\beta [ \theta^\prime(t) + \frac{1}{c} \sum_{\ell=1}^k J_\ell \sigma_\ell ] ) } \biggr\rangle_{\sigma_1,\cdots,\sigma_k} \label{eq:app_id1} \\
  \hspace*{-15mm}           
  &=& \sum_{k=0}^\infty p(k)
      \biggl( \prod_{\ell=1}^k  \int\! \rmd J_\ell ~\tilde{p}(J_\ell) \int\! \rmd h~ \mathcal{Q}(h) \biggr)
      \biggl( \prod_t \frac{C_{k,t} \rme^{\frac12 \beta E_{k,t}}} {\prod_{\ell=1}^k A_\ell} \biggr) \nonumber \\
  \hspace*{-15mm}      
  & & \times \prod_t
      \rme^{\beta \sigma(t+1) [ \theta(t) - \sum_{\ell=1}^k B_\ell + \frac 12 D_{k,t} + \frac 12 D_{k,t+2} + \frac 12 F_{k,t} \sigma(t-1) ]}
      \label{eq:app_id2} \\
  \hspace*{-15mm}     
  &=& \int\! \rmd h~
      \biggl\{
        \sum_{k=0}^\infty p(k)
        \biggl( \prod_{\ell=1}^k \int\! \rmd h_\ell \rmd J_\ell~ \tilde{p}(J_\ell) \mathcal{Q}(h_\ell) \biggr) 
        \prod_t \frac{\rme^{\beta \sigma(t+1)[h+\theta^\prime(t)]}}{2\cosh(\beta [h+\theta^\prime(t)])}
        \nonumber \\
  \hspace*{-15mm}       
  & &   \times \delta \biggl( h - \frac{2}{\beta c} \sum_{\ell=1}^k J_\ell \tanh \beta J_\ell + \frac 1{2\beta} \sum_{\ell=1}^k \ln \frac{\cosh(\beta[h_\ell+\frac{1}{c}J_\ell])}{\cosh(\beta[h_\ell-\frac{1}{c}J_\ell])} \biggr)
      \biggr\} \nonumber \\
  \hspace*{-15mm}  
  &=& \int\! \rmd h~
      \biggl\{
        \sum_{k=0}^\infty p(k)
        \biggl( \prod_{\ell=1}^k \int\! \rmd h_\ell\rmd J_\ell ~\tilde{p}(J_\ell) \mathcal{Q}(h_\ell) \biggr) 
        \prod_t \frac{e^{\beta \sigma(t+1)[h+\theta^\prime(t)]}}{2\cosh(\beta [h+\theta^\prime(t)])}
        \nonumber \\
  \hspace*{-15mm}        
  & &   \times \delta \biggl( h - \frac 1\beta \sum_{\ell=1}^k \tanh^{-1} [ (\tanh \beta h_\ell) (\tanh \frac{\beta J}{c}) ] \biggr)
      \biggr\}
\end{eqnarray}
where we have put
\begin{eqnarray}
  & &\hspace*{-20mm} A_\ell = \sqrt{ \cosh(\beta[h_\ell + \frac{1}{c} J_\ell]) \cosh(\beta[h_\ell - \frac{1}{c} J_\ell])}, \\
  & &\hspace*{-20mm} B_\ell = \frac 1{2\beta} \ln \frac{\cosh(\beta[h_\ell + \frac{1}{c}J_\ell])}{\cosh(\beta[h_\ell - \frac{1}{c}J_\ell])}, \\
  & &\hspace*{-20mm} C_{k,t} =
      \biggl\langle
        \frac
        {\exp(\beta \sum_{\ell=1}^k \sigma_\ell h_\ell)}
        {2 \cosh (\beta[ \theta^\prime(t)+\frac{1}{c}\sum_{\ell=1}^k  J_\ell \sigma_\ell])}
      \biggr\rangle_{\!\sigma_1,\cdots,\sigma_k} ,\\
  & &\hspace*{-20mm} D_{k,t} =
      \frac 1{2\beta} \ln
      \frac
        {
          \biggl\langle
          \frac
          {\exp(\beta \sum_{\ell=1}^k \sigma_\ell [h_\ell+\frac{2}{c}J_\ell])}
          {2 \cosh (\beta[ \theta^\prime(t)+\frac{1}{c}\sum_{\ell=1}^k J_\ell \sigma_\ell])}
          \biggr\rangle_{\!\sigma_1,\cdots,\sigma_k}
        }
        {
          \biggl\langle
          \frac
          {\exp(\beta \sum_{\ell=1}^k \sigma_\ell [h_\ell-\frac{2}{c}J_\ell])}
          {2 \cosh (\beta[ \theta^\prime(t)+\frac{1}{c}\sum_{\ell=1}^k J_\ell \sigma_\ell])}
          \biggr\rangle_{\!\sigma_1,\cdots,\sigma_k}
        }, \\
  & &\hspace*{-20mm} E_{k,t} =
      \frac 1{2\beta} \ln
      \frac
        {
          \biggl\langle
          \frac
          {\exp(\beta \sum_{\ell=1}^k \sigma_\ell [h_\ell+\frac{2}{c}J_\ell])}
          {2 \cosh (\beta[ \theta^\prime(t)+\frac{1}{c}\sum_{\ell=1}^k  J_\ell \sigma_\ell])}
          \biggr\rangle_{\!\sigma_1,\cdots,\sigma_k}
          \biggl\langle
          \frac
          {\exp(\beta \sum_{\ell=1}^k \sigma_\ell [h_\ell-\frac{2}{c}J_\ell])}
          {2 \cosh (\beta[ \theta^\prime(t)+\frac{1}{c}\sum_{\ell=1}^k  J_\ell \sigma_\ell])}
          \biggr\rangle_{\!\sigma_1,\cdots,\sigma_k}
        }
        {
          \biggl\langle
            \frac
            {\exp(\beta \sum_{\ell=1}^k \sigma_\ell h_\ell)}
            {2 \cosh (\beta[ \theta^\prime(t)+\frac{1}{c}\sum_{\ell=1}^k J_\ell \sigma_\ell])}
          \biggr\rangle_{\!\sigma_1,\cdots,\sigma_k}^2
        }. \nonumber\\&&
\end{eqnarray}
To obtain (\ref{eq:app_id1}) and (\ref{eq:app_id2}) we have used (\ref{eq:id1}) and (\ref{eq:id2}), respectively.
Equation (\ref{eq:ss_Q}) can be derived in the same way.


\begin{thebibliography}{99}




\bibitem{Viana1985}
L. Viana and A. J. Bray,
{\it J. Phys. C}, {\bf 18}, 3037, (1985).

\bibitem{Kanter1987}
I. Kanter and H. Sompolinsky,
{\it Phys. Rev. Lett.}, {\bf 58}, 164, (1987).

\bibitem{Mezard1987a}
M. Mezard and G. Parisi,
{\it Europhys. Lett.}, {\bf 3}, 1067, (1987).

\bibitem{Mezard1987b}
M. M\'ezard and G. Parisi, M. Virasoro,
{\it Spin glass theory and beyond},
World Scientific, (1987).

\bibitem{Mezard1987c}
M. M\'ezard and G. Parisi,
{\it Europhys. Lett.}, {\bf 3}, 1067 (1987).

\bibitem{Mottishaw1987}
P. Mottishaw and C. De Dominicis,
{\it J. Phys. A: Math. Gen.}, {\bf 20}, L375, (1987).

\bibitem{Wong1988}
K. Y. Wong and D. Sherrington D,
{\it J. Phys. A: Math. Gen.}, {\bf 21}, L459, (1988).

\bibitem{Monasson1988}
R. Monasson,
{\it J. Phys. A: Math. Gen.}, {\bf 31}, 513, (1988).

\bibitem{Coolen2005}
A. C. C. Coolen, N. S. Skantzos, I P\'erez Castillo, C. J. P\'erez Vicente, J. P. L. Hatchett, B. Wemmenhove and T. Nikoletopoulos,
{\it J. Phys. A: Math. Gen.}, {\bf 38}, 8289, (2005).

\bibitem{PerezVicente2008}
C. J. P\'erez Vicente and A. C. C. Coolen,
{\it J. Phys. A: Math. Theor.}, {\bf 41} 255003, (2008).


\bibitem{Sourlas1989}
N. Sourlas,
{\it Nature}, {\bf 339}, 693 (1989).

\bibitem{Kabashima1999}
Y. Kabashima and D. Saad,
{\it Europhys. Lett.}, {\bf 45}, 1, 97 (1999).

\bibitem{Vicente1999}
R. Vicente, D. Saad and Y. Kabashima,
{\it Phys. Rev. E}, {\bf 60}, 5, 5352 (1999).

\bibitem{Murayama2000}
T. Murayama, Y. Kabashima, D. Saad D and R. Vicente,
{\it Phys. Rev. E}, {\bf 62}, 1577, (2000).

\bibitem{Nakamura2001}
K. Nakamura, Y. Kabashima and D. Saad,
{\it Europhys. Lett.}, {\bf 56}, 610, (2001).

\bibitem{Nishimori2001}
H. Nishimori,
{\it Statistical Physics of Spin Glasses and Information Processing},
Oxford University Press, (2001).

\bibitem{Alamino2007}
R. C. Alamino and D. Saad,
{\it J. Phys. A: Math. Theor.}, {\bf 40}, 12259, (2007).

\bibitem{Neri2008}
I. Neri, N. S. Skantzos and D. Boll\'e,
{\it J. Stat. Mech.: Theory Exp.}, P10018, (2008).




\bibitem{Matsunaga2003}
Y. Matsunaga and H. Yamamoto,
{\it IEEE Trans. Inform. Theory}, vol. 49, 2225, (2003).

\bibitem{Murayama2003}
T. Murayama and M. Okada,
{\it J. Phys. A: Math. Gen.}, {\bf 36}, 11123, (2003).

\bibitem{Murayama2004}
T. Murayama,
{\it Phys. Rev. E}, vol. 69, 035105(R), 2004.

\bibitem{Wainwright2005}
M. J. Wainwright and E. Maneva,
{\it Proc. of 2005 Int'l. Sympo. Info. Theory (ISIT)}, 1493, 2005.

\bibitem{Ciliberti2005}
S. Ciliberti, M. M\'ezard,
{\it J Stat. Mech.}, vol. 3, 58, (2006).

\bibitem{Ciliberti2006}
S. Ciliberti, M. M\'ezard and R. Zecchina,
{\it Complex systems methods}, vol. 3, 58, (2006).

\bibitem{Martinian2006}
E. Martinian and M. J. Wainwright,
{\it Workshop on Info. Theory and its Applications}, (2006).

\bibitem{Mimura2009a}
K. Mimura,
{\it J Phys. A: Math. Theor.}, {\bf 42}, 135002, (2009).




\bibitem{Yoshida2005}
M. Yoshida and T. Tanaka,
{\it Proc. of 2006 Int'l Sympo. Info. Theory (ISIT)}, 2378, (2006).

\bibitem{Raymond2007}
J. Raymond and D. Saad,
{\it J Phys. A: Math. Theor.}, {\bf 40}, 12315, (2007).

\bibitem{Guo2008}
D. Guo and C. Wang,
{\it IEEE Journal on Selected Areas in Comm.}, {\bf 26}, 421, (2008).




\bibitem{Kirkpatric1994}
S. Kirkpatrick and B. Selman,
{\it Science}, {\bf 264}, 1297, (1994).

\bibitem{Monasson1998a}
R. Monasson and R. Zecchina,
{\it Phys. Rev. E}, {\bf 56}, 1357, (1998).

\bibitem{Monasson1998b}
R. Monasson and R. Zecchina,
{\it J. Phys. A: Math. Gen.}, {\bf 31}, 9209, (1998).

\bibitem{Monasson1999}
R. Monasson, R. Zecchina, S. Kirkpatrick, B. Selman and L. Troyansky,
{\it Nature}, {\bf 400}, 133, (1999).

\bibitem{Zdeborova2008}
L. Zdeborov\'a and Marc M\'ezard,
{\it Phys. Rev. Lett.}, {\bf 101}, 078702, (2008).




\bibitem{Castillo2004a}
I. P\'erez Castillo and N. S. Skantzos,
{\it J. Phys. A: Math. Gen.}, {\bf 37}, 9087, (2004).

\bibitem{Castillo2004b}
I. P\'erez Castillo, B. Wemmenhove, J. P. L. Hatchett, A. C. C. Coolen, N. S. Skantzos, and T. Nikoletopoulos,
{\it J. Phys. A: Math. Gen.}, {\bf 37}, 8789, (2004).




\bibitem{Gitterman2000}
A. Gitterman,
{\it J. Phys. A: Math. Gen.}, {\bf 33}, 8373, (2000).

\bibitem{Nikoletopoulos2004}
T. Nikoletopoulos, A. C. C. Coolen, I. P\'erez Castillo, N. S. Skantzos, J. P. L. Hatchett and B. Wemmenhove,
{\it J. Phys. A: Math. Gen.}, {\bf 37}, 6455, (2004).

\bibitem{Hatchett2005b}
J. P. L. Hatchett, N. S. Skantzos, and T. Nikoletopoulos,
Phys. Rev. E 72, 066105 (2005).

\bibitem{Skantzos2005}
N. S. Skantzos, Isaac P\'erez Castillo, and J. P. L. Hatchett,
Phys. Rev. E 72, 066127 (2005).




\bibitem{Semerjian2003}
G. Semerjian and L. F. Cugliandolo,
{\it Europhys. Lett.}, {\bf 61}, 2, 247 (2003).

\bibitem{Semerjian2004a}
G. Semerjian G, L. F. Cugliandolo and A. Montanari,
{\it J. Stat. Phys.}, {\bf 115}, 493, (2004).

\bibitem{Semerjian2004b}
G. Semerjian G and M. Weigt,
{\it J. Phys. A: Math. Gen.}, {\bf 37}, 5525, (2004).

\bibitem{Hatchett2004}
J. P. L. Hatchett, B. Wemmenhove, I. P\'erez Castillo, T. Nikoletopoulos, N. S. Skantzos and A. C. C. Coolen,
{\it J. Phys. A: Math. Gen.}, {\bf 37}, 6201, (2004).

\bibitem{Hatchett2005a}
J. P. L. Hatchett, I. P\'erez Castillo, A. C. C. Coolen, and N. S. Skantzos,
{\it Phys. Rev. Lett.}, {\bf 95}, 117204, (2005).

\bibitem{Skantzos2007}
N. S. Skantzos and J. P. L. Hatchett,
{\it Physica A}, {\bf 381}, 202, (2007).

\bibitem{Mozeika2008}
A. Mozeika and A. C. C. Coolen,
{\it J. Phys. A: Math. Theor.}, {\bf 41}, 115003, (2008).

\bibitem{Mozeika2009}
A. Mozeika and A. C. C. Coolen,
{\it J. Phys. A: Math. Theor.}, {\bf 42}, 195006, (2009).

\bibitem{Mimura2009b}
K. Mimura and A. C. C. Coolen,
{\it Proc. of 2009 Int'l Sympo. Info. Theory (ISIT)}, 1829, (2009).

\bibitem{Neri2009}
I. Neri and D. Boll\'e, "The cavity approach to the parallel dynamics on finitely connected graphs",
arXiv:0905.3260.




\bibitem{Dominicis1978}
C. De Dominicis,
{\it Phys. Rev. B}, {\bf 18}, 4913, (1978).



\bibitem{Peretto1984}
P. Peretto,
{\it Biol. Cybern.}, {\bf 50}, 51 (1984).




\end{thebibliography}
\end{document}